\documentclass[twocolumn]{aastex62}

\newcommand\revision[1]{\textcolor{black}{#1}}
\newcommand{\s}[1]{\textsubscript{#1}}
\newcommand{\Msun}{ M\(_\odot\)}

\usepackage{xcolor}
\usepackage{amsmath,amssymb,gensymb,times,graphicx,morefloats,color}
\usepackage{amsmath}
\usepackage{graphics,graphicx,url,verbatim,xspace}

\usepackage{changepage}
\usepackage{mathtools} 
\usepackage{comment}
\usepackage{gensymb}

\graphicspath{{./}{figures/}}

\accepted{\today}
\submitjournal{ApJ}

\shorttitle{Boyajian's Star B: The co-moving stellar companion to KIC~8462852}
\shortauthors{Pearce et al.}

\begin{document}
\defcitealias{boyajian_planet_2016}{B16}
\defcitealias{deacon_pan-starrs_2016}{D16}
\title{Boyajian's Star B: The co-moving companion to KIC~8462852 A}

\correspondingauthor{Logan A. Pearce}
\email{loganpearce1@email.arizona.edu}

\author[0000-0003-3904-7378]{Logan A. Pearce}
\affil{Steward Observatory, University of Arizona, Tucson, AZ 85721, USA}
\affil{NSF Graduate Research Fellow}

\author[0000-0001-9811-568X]{Adam L. Kraus}
\affil{Department of Astronomy, University of Texas at Austin, Austin, TX, 78712, USA}

\author[0000-0001-9823-1445]{Trent J. Dupuy}
\affil{Institute for Astronomy, University of Edinburgh, Blackford Hill, Edinburgh, EH9 3HJ, United Kingdom}

\author[0000-0003-3654-1602]{Andrew W. Mann}
\affiliation{Department of Physics and Astronomy, The University of North Carolina at Chapel Hill, Chapel Hill, NC 27599, USA} 

\author[0000-0001-8832-4488]{Daniel Huber}
\affiliation{Institute for Astronomy, University of Hawai`i, 2680 Woodlawn Drive, Honolulu, HI 96822, USA} 

\begin{abstract}
The light curve of KIC~8462852, a.k.a Boyajian's Star, undergoes deep dips the origin of which remains unclear.  A faint star $\approx$2\arcsec to the east was discovered in Keck/NIRC2 imaging in Boyajian et al. (2016), but its status as a binary, and possible contribution to the observed variability, was unclear.  Here, we use three epochs of Keck/NIRC2 imaging, spanning five years, in \textsl{JHK} near-infrared bands to obtain 1-mas precision astrometry.  We show that the two objects exhibit common proper motion, measure a relative velocity of $\mu=0.14\pm0.44$\,mas\,yr$^{-1}$ ($\mu=0.30\pm0.93$\,km\,s$^{-1}$) and conclude that they are a binary pair at $880\pm10$~AU projected separation.  
\revision{There is marginal detection of possible orbital motion, but our astrometry is insufficient to characterize the orbit.}
We show that two other point sources are not associated with KIC~8462852. 
We recommend that attempts to model KIC~8462852~A's light curve should revisit the possibility that the bound stellar companion may play a role in causing the irregular brightness variations, for example through disruption of the orbits of bodies around the primary due to long-term orbital evolution of the binary orbit.

\end{abstract}


\section{Introduction}\label{sec:intro}

Many possible causes have been posited to explain the unusual light curve of KIC~8462852 (a.k.a ``Boyajian's Star'') \revision{discovered in \citealt[][]{boyajian_planet_2016} (hereafter B16)}. KIC~8462852 exhibits large, possibly aperiodic dips in a variety of shapes, inconsistent with an exoplanet explanation \citepalias{boyajian_planet_2016}.  There has been significant interest in the system, with many suggested explanations from the beginning of a Late Heavy Bombardment-like period \citep{bodman_kic_2016}, to interstellar clouds or an intervening object \citep{wright_families_2016}, to uneven circumstellar material \citep{wyatt_modelling_2018}, to alien megastructures \citep{wright_search_2016}. Some explanations such as recent cataclysmic dust-generating events \citep{marengo_kic_2015}, massive debris disks \citep{thompson_constraints_2016}, close-in obscuring material or YSO-like behavior \citep{lisse_irtfspex_2015}, and instrumental effects \citepalias{boyajian_planet_2016} have already been ruled out as explanations.  

The break-up of exocomets or planetesimals on eccentric orbits was preferred given the observations \citep{thompson_constraints_2016}, although \citet{bodman_kic_2016} showed this idea does not fully explain all the dips, nor the apparent long term dimming trend \citep{montet_kic_2016}.  \citet{simon_where_2018} showed the long-term dimming trend may be part of a more complicated episodic dimming and brightening.  \citet{boyajian_first_2018} report post-\textsl{Kepler} observations that show consistency with optically thin dust and intrinsic variations of the star, while 
\citet{martinez_highres_notzotero} found no clear evidence of comets and evidence for clumps of thick material within the thin dust. 

A faint possible companion star was observed at separation $\rho = 1.95\arcsec$, position angle PA$ = 96\fdg6$, and $\Delta{H}= 3.840\pm0.017$\,mag by \citetalias{boyajian_planet_2016}, but given their single epoch of imaging, were unable to determine if the two were physically associated.  \citetalias{boyajian_planet_2016} determined that blending with the object was not a cause of the anomalous light curve, as its optical faintness means even a 100\% drop in its flux could not explain the deepest dips seen in the Kepler light curve.  
The object's separation translates to $880\pm10$\,AU at the distance of $451\pm5$\,pc \citep{gaia_collaboration_gaia_2018}, meaning it would not be currently affecting the observed behavior of KIC~8462852 via tides or strong gravitational interactions with bodies at small orbital radii.  If it were a bound companion however, it might affect other bound objects via long-term perturbations, and could trigger a barrage of occulting objects inward towards the host star.

\citealt{clemens_proper_2018} observed the system with the Mimir near-infrared wide-field imager on the 1.8-m Perkins telescope in 2017, and compared the relative astrometry of the candidate companion to the 2014 Keck/NIRC2 observations reported in \citetalias{boyajian_planet_2016}.  They reported a tangential speed of 44.9$\pm$4.9 km s$^{-1}$ for the candidate companion relative to KIC~8462852, and concluded it is not a bound companion. However, the seeing-limited resolution of their second-epoch imaging was 1.3--1.5\arcsec, so the companion was only resolved at $\le$1.5 times the observational FWHM. This challenging observational fit could have been prone to a separation/contrast degeneracy that affected the measured change in position.

In this work we use three epochs of Keck/NIRC2 astrometry spanning five years to revisit the status of the close companion to KIC~8462852, and \revision{show that they are a common proper motion pair and a gravitationally bound binary system.}
We analyze two other faint objects in our images and show that they are unassociated.  
In Section \ref{sec:analysis} we outline our astrometric methods.  In Section \ref{sec:results} we report relative proper motions and demonstrate common (or lack of common) proper motion for the three candidate companions from our astrometry and \revision{assess the probability of binarity}.  In Section \ref{sec:discussion} we discuss implications of the presence of a wide stellar companion for the KIC~8462852 system.

\section{Analysis}\label{sec:analysis}
\subsection{Observations}


We obtained observations of the KIC~8462852 system using the near-infrared imaging camera NIRC2 coupled with the natural guide star adaptive optics system \citep{Wizinowich2000} on the Keck-II telescope in 2014 (28 images), 2016 (13 images), and 2019 (10 images). The observations in 2014 (PI M. Liu) were obtained in $K$, $H$, and $J$ bands, and were used by \citetalias{boyajian_planet_2016} to confirm the existence of KIC~8462852~B as a candidate companion. 
The observations in 2016 and 2019 were obtained by our team (PIs Mann and Huber) in the $K^{\prime}$-band filter, with the aim of testing whether KIC~8462852~B is co-moving. In both 2016 and 2019, we obtained observations with KIC~8462852 placed at two distinct orientations and dither positions. We deliberately duplicated the 2016 orientations and positions in 2019, with the goal of enabling cross-epoch measurements of motion that are independent of residual errors in the correction of the camera's static geometric distortion. In our subsequent analysis, we henceforth treat each of the two dither positions in 2016 and 2019 as an independent observation, denoted as 2016-1/2019-1 and 2016-2/2019-2, respectively.  

All images used the narrow camera, with adaptive optics in natural guide star mode, in position angle tracking mode.   We linearized each science and calibration frame in Python using the methodology of the IDL task \texttt{linearize\textunderscore nirc2.pro} \footnote{http://www.astro.sunysb.edu/metchev/ao.html} \citep{Metchev2009}, then dark-subtracted and flat-fielded science frames in the standard manner.  We adopted bad pixel identifications from \cite{stellar} and replaced with the median of surrounding pixels, then detrended spatially correlated readnoise from the mirrored positions of each quadrant (Kraus et al., in prep). 

\begin{figure*}
\centering
\includegraphics[width=0.9\textwidth]{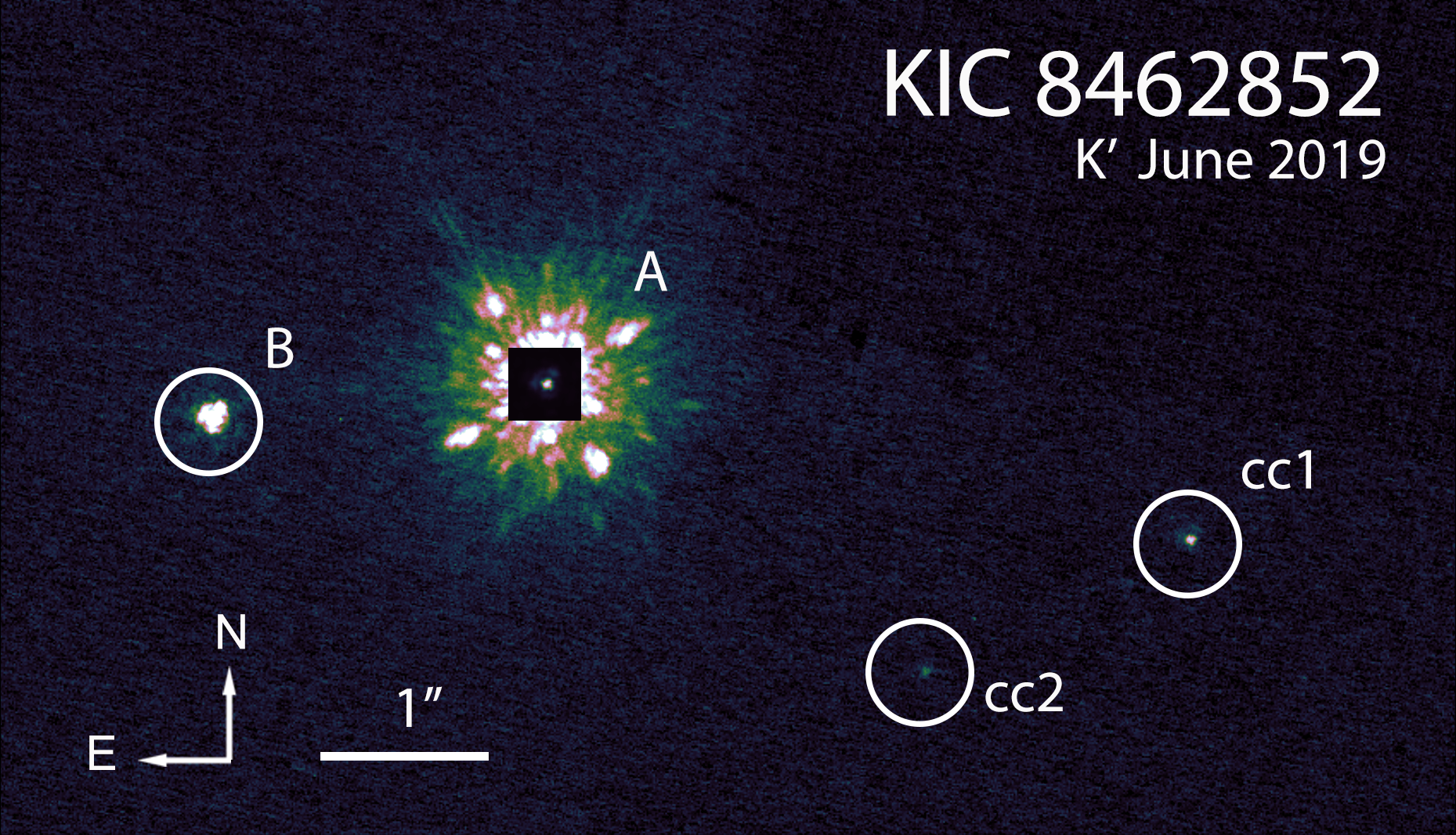}
\caption{\small{Keck/NIRC2 adaptive optics image of KIC~8462852 A, B, and candidate companions, shown in log stretch to emphasize the faint candidate companions.  The primary, KIC~8462852 A, is shown inside a linear stretched box to avoid saturation.  The secondary, labeled B, is located 2\arcsec\, to the east.  The two candidate companions are labeled cc1 for the brighter companion, 3.8\arcsec\, southwest, and cc2, 2.8\arcsec\, southwest. }}
\label{fig:image}
\end{figure*}

Figure \ref{fig:image} displays a NIRC2 image from the 2019 epoch, with the comoving binary companion marked as B, located 1.95\arcsec\ to the east, and the two candidate companions marked as cc1, 3.8\arcsec\ southwest, and cc2, 2.8\arcsec\ southwest.  KIC~8462852~B was visible at sufficient signal-to-noise for astrometric analysis in all datasets; cc1 was sufficiently visible only in the $K$ and $K^{\prime}$ bands at all three epochs; and cc2 was sufficiently visible only in the $K^{\prime}$ band in the 2016 and 2019 datasets.

\floattable
\begin{deluxetable}{lcccccccc}
\tablecaption{{Keck/NIRC2 NGS AO Astrometry for KIC~8462852 candidate companions}\label{table:summary}}
\tablehead{
\colhead{Epoch} & \colhead{MJD} & \colhead{Filter} & \colhead{N$_{\rm images}$} & \colhead{N$_{\rm coadds}$} & \colhead{t$_{int}$} & \colhead{Separation\tablenotemark{a}} & \colhead{Position Angle\tablenotemark{b}} & \colhead{$\Delta{m}\tablenotemark{c}$} \\
\colhead{} & \colhead{} & \colhead{} & \colhead{}& \colhead{} & \colhead{(sec)} & \colhead{(mas)} & \colhead{(deg)} & \colhead{(mag)} 
}
\startdata
\multicolumn{9}{c}{B} \\
\hline
2014.79 & 56946.30 & $J$ & 9 & 10 & 0.726  & 1952.78 $\pm \,0.77$ & 96.064 $\pm \,0.014$ & 3.884 $\pm$ 0.057 \\
2014.79 & 56946.30 & $H$ & 9 & 10 & 0.726  & 1952.69 $\pm \,0.36$ & 96.059 $\pm \,0.010$ & 3.704 $\pm$ 0.034 \\
2014.79 & 56946.30 & $K$ & 10 & 10 & 0.726  & 1952.61 $\pm \,0.40$ & 96.058 $\pm \,0.013$ & 3.525 $\pm$ 0.020 \\
2016.72 Dither 1 & 57651.23 & $K^\prime$ & 2 & 10 & 1.0 & 1950.64 $\pm \,0.14$ & 96.064 $\pm \,0.004$ & 3.640 $\pm$ 0.004 \\
2016.72 Dither 2 & 57651.23 & $K^\prime$ & 11 & 10 & 1.0 & 1951.07 $\pm \,0.07$ & 96.063 $\pm \,0.004$ & 3.638 $\pm$ 0.012 \\
2019.44 Dither 1 & 58646.50 & $K^\prime$ & 2 & 20 & 1.0 & 1951.63 $\pm \,0.09$ & 96.069 $\pm \,0.004$ & 3.632 $\pm$ 0.006 \\
2019.44 Dither 2 & 58646.50 & $K^\prime$ & 8 & 20 & 1.0 & 1951.88 $\pm \,0.06$ & 96.062 $\pm \,0.004$ & 3.626 $\pm$ 0.009 \\
\hline
\multicolumn{9}{c}{cc1} \\
\hline
2014.79 & 56946.30 & $K$ & 10 & 10 & 0.726  & 3871.5 $\pm\,2.6$ & 256.377 $\pm\,0.042$ & 5.77 $\pm$ 0.38\\
2016.72 Dither 1 & 57651.23 & $K^\prime$ & 2 & 10 & 1.0 & 3863.2 $\pm \,1.3$ & 256.380 $\pm \,0.015$ & 6.11 $\pm$ 0.03 \\
2016.72 Dither 2 & 57651.23 & $K^\prime$ & 11 & 10 & 1.0 & 3862.0 $\pm \,0.7$ & 256.401 $\pm \,0.007$ & 6.07 $\pm$ 0.05 \\
2019.44 Dither 1 & 58646.50 & $K^\prime$ & 2 & 20 & 1.0 & 3850.1 $\pm \,1.1$ & 256.370 $\pm \,0.012$ & 6.23 $\pm$ 0.02 \\
2019.44 Dither 2 & 58646.50 & $K^\prime$ & 8 & 20 & 1.0 & 3848.3 $\pm \,0.5$ & 256.367 $\pm \,0.008$ & 6.24 $\pm$ 0.04 \\
\hline
\multicolumn{9}{c}{cc2} \\
\hline
2016.72 Dither 1 & 57651.23 & $K^\prime$ & 2 & 10 & 1.0 & 2785 $\pm \,7$ & 232.51 $\pm \,0.02$ & 7.15 $\pm$ 0.09 \\
2016.72 Dither 2 & 57651.23 & $K^\prime$ & 5 & 10 & 1.0 & 2788 $\pm \,6$ & 232.46 $\pm \,0.05$ & 7.04 $\pm$ 0.11 \\
2019.44 Dither 1 & 58646.50 & $K^\prime$ & 2 & 20 & 1.0 & 2761 $\pm \,3$ & 232.49 $\pm \,0.03$ & 7.26 $\pm$ 0.10 \\
2019.44 Dither 2 & 58646.50 & $K^\prime$ & 7 & 20 & 1.0 & 2763 $\pm \,2$ & 232.52 $\pm \,0.03$ & 7.44 $\pm$ 0.13 \\
\enddata
\tablenotetext{a}{Errors shown are the statistical error, corresponding to thick error bars on Figure \ref{fig:astr_B}, \ref{fig:astr_cc1}, and \ref{fig:astr_cc2}.  Systematic error of 1.4 mas applies to all separation measurements, corresponding to thin error bars.}
\tablenotetext{b}{Errors shown are the statistical error.  A systematic error applies to all position angle measurements which corresponds to 1.4 mas tangential angular distance at the object's separation:  0.042$\degree$ for B, 0.021$\degree$ for cc1, 0.029$\degree$ for cc2.}
\tablenotetext{c}{Errors shown are the statistical error. We adopt a conservative systematic error of 0.05 mag to account for detector systematics.}
\tablecomments{ 
0.029 deg = 1 mas tangential angular distance at the projected separation of B.   
N\s{images} is the number of images in each epoch.  t\s{int} is integration time per coadd.   Some images were excluded from the astrometry of cc2 due to insufficient detection of the object in those images.}  
\end{deluxetable}

\subsection{Astrometry}
We used the Gaussian PSF fitting routine described in \citet{Pearce2019GSC6214-210} to precisely measure the $(x,y)$ pixel position and uncertainty of the four components in each image.  Details of the modeling and acceptance criterion are described in \citet{Pearce2019GSC6214-210} and applied in the same manner to this data set.  Briefly, we modeled the PSF of the primary and candidate companion as the sum of two 2-dimensional Gaussian functions and varied the model parameters through a custom Gibbs Sampler Markov Chain Monte Carlo (MCMC) routine. 
We performed astrometric calibrations for the primary and each of the three candidate companions for each step along the MCMC chains, including optical distortion and plate scale error \citep{yelda_improving_2010, service_new_2016}, and differential aberration and atmospheric refraction, then computed separation and position angle from primary for each $(x,y)$ position in the chains for each candidate companion.  We took the mean and standard deviation of the separation and position angle chains as the final value and uncertainty in an image.  
Positions in each image are given in Figures \ref{fig:astr_B}, \ref{fig:astr_cc1}, and \ref{fig:astr_cc2} as filled circles, with the median error for images in an epoch given by offset crosses. We computed a weighted mean of image positions as the mean position in an epoch, given by small crosses in those figures.

The geometric distortion solution for NIRC2 removes almost all of the distortion introduced by the telescope and instrument, but it does leave residual systematic uncertainties of $\sigma_{s,pos} \sim$1\,mas in the position measurements of individual sources, or $\sigma_{s} \sim \sqrt{2}$\,mas in the relative astrometry between two sources. For a given detector location, any further residual error associated with that location (such as from temperature variations and non-repeatable positioning of the telescope/instrument optics) appears to be negligible across a timescale of years when compared to the empirical scatter seen within an individual epoch ($\la$0.2\,mas;  \citealt{dupuy_orbital_2016}; Dupuy et al., in prep) 
It is therefore possible to achieve substantial further improvement in the precision of relative astrometric measurements if source positions and orientations can be duplicated in multiple epochs. Our observations in 2019 replicated the two position/orientation arrangements used in 2016, so we have encapsulated these systematic uncertainties in the covariance matrix for the five epochs (2014, 2016-1, 2019-1, 2016-2, 2019-2):

\begin{equation*}
C = 
\begin{bmatrix}
\sigma_{14}^2 + \sigma_{s}^2 & 0 & 0 & 0 & 0 \\
0 & \sigma_{16{\text -}1}^2 + \sigma_{s}^2 & \sigma_{s}^2 & 0 & 0 \\
0 & \sigma_{s}^2 & \sigma_{19{\text -}1}^2 + \sigma_{s}^2 & 0 & 0 \\
0 & 0 & 0 & \sigma_{16{\text -}2}^2 + \sigma_{s}^2 & \sigma_{s}^2 \\
0 & 0 & 0 & \sigma_{s}^2 & \sigma_{19{\text -}2}^2 + \sigma_{s}^2 
\end{bmatrix}
\end{equation*}

\noindent
where each diagonal term contains a contribution from the statistical variance $\sigma_{NN}^2$ (estimated from the RMS of the individual-image measurements at that epoch) as well as the systematic variance $\sigma_s^2$, and the systematic variance also contributes to the off-diagonal terms for epochs that were taken with the same source positions/orientations and hence share a common systematic error. 

In subsequent measurements of the $\chi^2$ goodness-of-fit statistic, we then use its modified definition:

\begin{equation*}
\chi^2 = \textbf{r}^T \, \textbf{C}^{-1} \, \textbf{r}
\end{equation*}

\noindent
where $r$ is the vector of residuals for the observations about the model being tested and $C^{-1}$ is the weight matrix corresponding to the inverse of the observational covariance matrix.  

We computed a relative velocity using the \texttt{scipy} least squares fitting function \texttt{curve\_fit} \citep{virtanen_scipy_2020} by fitting a linear function to the mean positions, weighted with the weight matrix $C^{-1}$.  We also use this formulation of the $\chi^2$ goodness-of-fit in the orbit fitting calculations presented in Section \ref{sec:orbit}.   We computed the contrast in each filter as the mean and standard deviation of the flux ratio of our analytical models in each image. 

Table \ref{table:summary} displays the results of our relative astrometry and photometry for the three candidate companions. 


\begin{figure}
\centering
\includegraphics[width=0.45\textwidth]{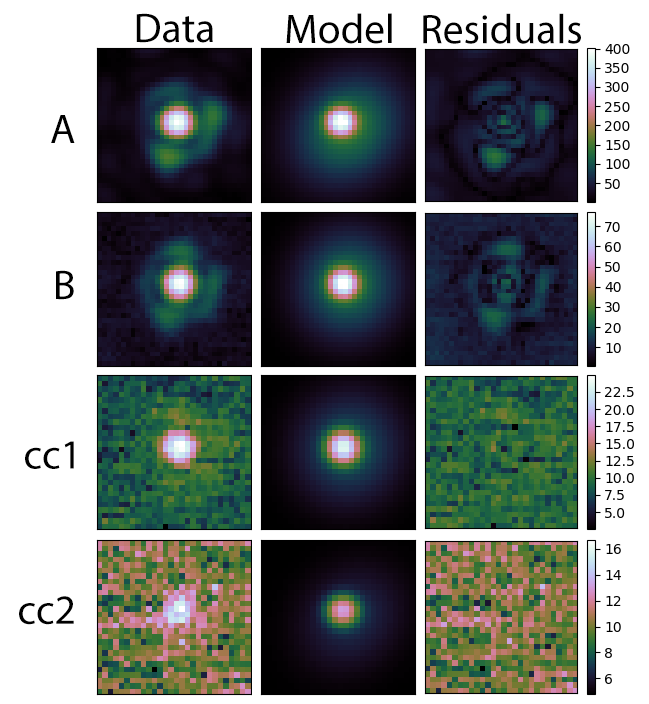}
\caption{\small{Data, model, and residual map of the primary (labeled "A"), companion (labeled "B"), and candidate companions ("cc1" and "cc2") for one image from the 2019 dataset.  The model shown is built using the mean values of the parameter chains from the MCMC fit for that image, and is shown with a square root stretch to emphasize the faint residuals. }}
\label{fig:residuals}
\end{figure}

\subsection{Stellar parameters}

In Table~\ref{table:stellar parameters} we summarize relevant properties of KIC~8462852~AB.  \citetalias{boyajian_planet_2016} found that the primary (F3V, $M = 1.43$\,\Msun; $T_{\rm{eff}} = 6750\pm120$\,K) has a space velocity inconsistent with any moving groups, and were unable to estimate an age.

The nearby companion, KIC~8462852~B, is comoving with A (demonstrated in Section \ref{sec:results B}), so its stellar parameters are also relevant to establish. In the absence of spectroscopic followup, we adopt the spectral type of M2V estimated by \citetalias{boyajian_planet_2016}. 

To update the stellar parameters for both components we used \texttt{isoclassify} \citep{huber17}, with input constraints including T$_{\rm{eff}}$ for the primary from \citetalias{boyajian_planet_2016}, the 2MASS K magnitude, Gaia DR2 parallax \citep[corrected for the zeropoint offset in the Kepler field,][]{zinn19}, a solar metallicity prior with a width of 0.1 dex and the median measured $K^\prime$ contrast in Table~\ref{table:summary} with a conservative uncertainty accounting for errors in the synthetic fluxes ($\Delta K^\prime=3.64\pm0.05$ mag). This procedure is essentially the same as in \citet{kraus_impact_2016}, but with an improved stellar classification method and a newer grid of MIST isochrones \citep{choi16} supplemented with the empirical relations by \citet{mann15} and \citet{mann_how_2019} for low-mass stars, as described in \citet{berger20}. The resulting classification yields a self-consistent classification of the primary and secondary assuming both components have the same age and metallicity. We estimate an isochronal age of $\sim$1.2 Gyr, and the resulting updated stellar parameters are listed in Table~\ref{table:stellar parameters}. Note that the uncertainties do not account for systematic errors between different model grids. 


\begin{deluxetable}{ccc}[htb!]
\tablecaption{{System and Stellar Properties for KIC~8462852~AB}\label{table:stellar parameters}}
\tablehead{\colhead{Property}  & \colhead{} & \colhead{Ref} }
\startdata
\hline
Distance (pc) & 451.0$^{+4.9}_{-4.8}$ & 1 \\
$\rho$ (mas) & $1951.48 \pm 0.23$ & \revision{Sec 3.1} \\
$\rho$ (AU) & $880 \pm 10$ & 2, \revision{Sec 3.1} \\
$PA$ ($^{\circ}$) & $96.063 \pm 0.004$ & \revision{Sec 3.1} \\
\hline
\multicolumn{3}{c}{\it KIC~8462852 A}\\
Proper Motion  & $\mu_{\alpha}$=-10.422 $\pm$ 0.040 & 2\\
(mas yr\textsuperscript{-1}) &$\mu_{\delta}$=-10.288 $\pm$ 0.041 & 2 \\
Luminosity (L\(_\odot\)) & 4.3 $\pm$ 0.3 & \revision{Sec 2.3}\\
Mass (\Msun) & 1.36 $\pm$ 0.05 & \revision{Sec 2.3}\\
Radius (R\(_\odot\)) & 1.51 $\pm$ 0.04 & \revision{Sec 2.2}\\
T$_{\rm{eff}}$ (K) & 6750 $\pm$ 120 & 3\\
SpT & F3V & 3\\
Age (Gyr) & $\sim$1.2 & \revision{Sec 2.3}\\
\revision{$J$ (mag)} & 10.763 $\pm$ 0.021 & 4 \\
\revision{$H$ (mag)} & 10.551 $\pm$ 0.019 & 4 \\
\revision{$K$ (mag)} & 10.499 $\pm$ 0.020 & 4 \\
\hline
\multicolumn{3}{c}{\it KIC~8462852~B}\\
Mass (\Msun) & $0.44\pm0.02$ & 5, \revision{Sec 2.3}\\
Radius (R\(_\odot\)) & 0.45 $\pm$ 0.02 & \revision{Sec 2.3}\\
T$_{\rm{eff}}$ (K) & 3720 $\pm$ 70 & \revision{Sec 2.3} \\
SpT & M2V & 3\\
\enddata
\tablereferences{\small{(1) \citet{bailer-jones_estimating_2018} (2) {\it Gaia} DR2 \citep{gaia_collaboration_gaia_2016, gaia_collaboration_gaia_2018}; (3) \citet{boyajian_planet_2016}; \revision{(4) 2MASS \citep{skrutskie_two_2006}}; (5) \citet{mann_how_2019}}}
\end{deluxetable}

\section{Results}\label{sec:results}
\subsection{KIC~8462852~B}\label{sec:results B}
\begin{figure*}
\centering
\includegraphics[width=0.55\textwidth]{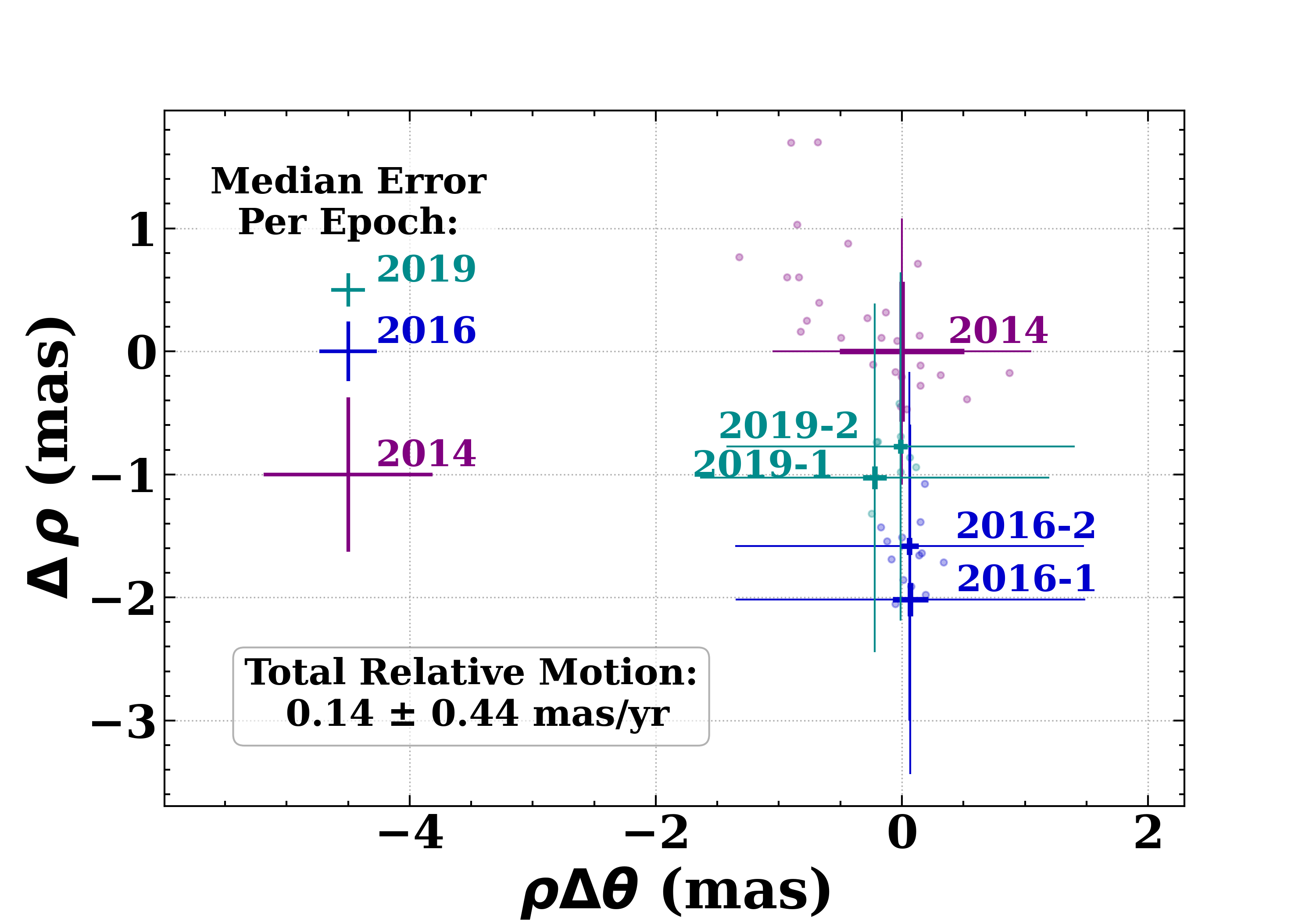}
\includegraphics[width=0.35\textwidth]{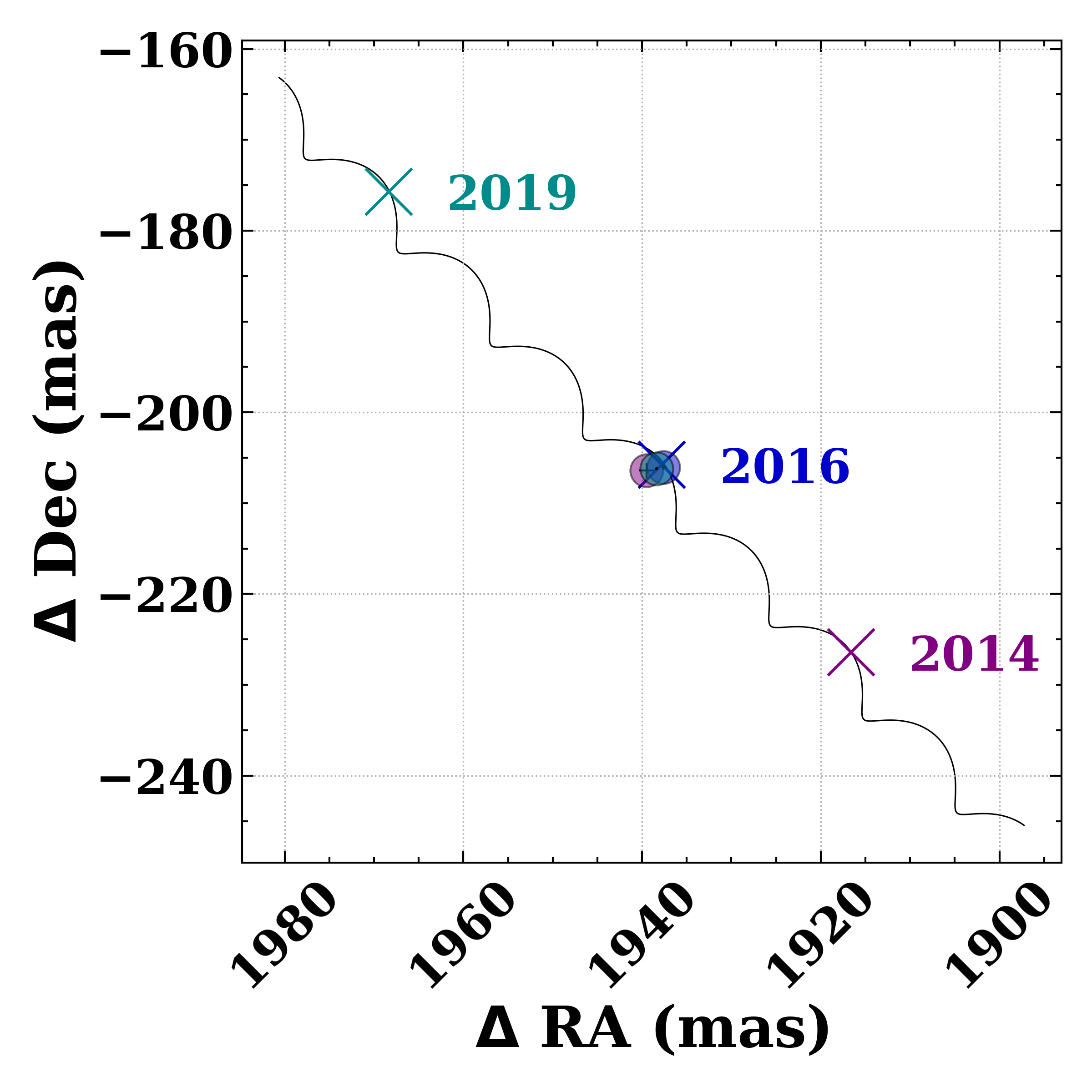}
\caption{\small{Left: Change in relative astrometry for KIC~8462852~B in separation (y-axis) and angular direction (x-axis) 
in individual images (circles) and mean values for each epoch (crosses).  Epochs 2016 and 2019 dither positions are reported as separate observations. Thick crosses show statistical uncertainty, thin crosses show systematic uncertainty.  Crosses to the left display the median error in individual image measurements.  We measure a total relative velocity of $\mu = 0.14 \pm 0.44$ mas yr$^{-1}$, which is consistent with zero.  Right: Observed position of KIC~8462852~B (circles, error bars smaller than marker size) with expected motion if it were a background star (black track, crosses indicate expected position at observation times).  KIC~8462852~B displays common proper motion with KIC~8462852 A, consistent with a bound companion. }}
\label{fig:astr_B}
\end{figure*}

\revision{Our measured relative motion indicates that KIC~8462852 and its close, bright neighbor are a common proper motion pair.}
We determined a relative motion in the plane of the sky of $\Delta\mu=0.14\pm0.44$\,mas\,yr$^{-1}$ ($\Delta\mu=0.3\pm1.0$\,km\,s$^{-1}$) over the five year span of observations, which is consistent with a bound companion.  Figure \ref{fig:astr_B} (left) displays the motion of B relative to the primary.  As discussed above, the 2016 epoch was obtained in two dither positions; we matched observations in 2019 to the same two positions on the detector as 2016, to enable comparison.  The change in position from 2016 to 2019 is consistent between images in the same dither position (i.e. 2016-1 to 2019-1 is consistent with 2016-2 to 2019-2).  Our astrometry is sufficiently precise that the differences between 2016 and 2019 could be due to orbital motion. 
The 2014 epoch is offset from the others, most likely due to the NIRC2 realignment which occurred in 2015 \citep{service_new_2016}. 

\revision{KIC~8462852~A and B are in {\it Gaia} EDR3\footnote{Source IDs: A: 2081900940499099136, B: 2081900944807842560, cc1: 2081900944800715648} \citep{GaiaEDR3summary,gaia_collaboration_gaia_2016}, which was released to the public while this paper was under review.  {\it Gaia} EDR3 reports a total relative proper motion within 1-$\sigma$ of our measured value ($0.44 \pm 0.34$ mas yr$^{-1}$) and parallaxes for A and B that are consistent to within their uncertainty. We continued to use {\it Gaia} DR2 for some calculations such as stellar parameters due to independent validations and the Kepler field zero point.}

\revision{\subsubsection{Tests for binarity}\label{sec: binarity}}
To test the level of consistency with co-movement, we computed the $\chi^2$ goodness-of-fit for the specific cases of the companion being a distant background object (zero absolute proper motion) or completely co-moving (zero relative proper motion).
Figure \ref{fig:astr_B} (right) displays the common proper motion of KIC~8462852~B with KIC~8462852~A.  Our measurements reject the null hypothesis that the object is a non-moving background star with zero proper motion ($\chi^2_{\rm{non-moving}} = 1060$, $\chi^2_{\rm{co-moving}} = 25.6$, for 4 degrees of freedom).  We interpret the disagreement with the zero relative proper motion case as likely due to orbital motion.

\revision{We performed two additional statistical tests to assess the probability of observing a non-bound star at the position, velocity, magnitude, and parallax of the common proper motion candidate companion.  First, we used a statistical approach similar to Sec. 4.5 of \citep{Correia2006VisualPMSBinaries} to estimate the probability of chance alignment given the surface density of similar objects in the vicinity.  We queried the {\it Gaia} EDR3 catalog for objects within a 30$\degree$ radius of KIC~8462852~A with proper motion within 1-$\sigma$ of the {\it Gaia} proper motion of KIC~8462852~A ($\mu_\alpha = 10.4 \pm 0.6$ mas yr$^{-1}$, $\mu_\delta = 10.3 \pm 0.6$ mas yr$^{-1}$) and parallax within 1-$\sigma$ of KIC~8462852~A ($\pm 0.025$ mas), in order to determine the most conservative comparison. This returned 140 objects, a surface density of $\Sigma = 3.8\times10^{-9}$ arcsec$^{-2}$.  The probability of observing a field object within $\theta$ = 2$\arcsec$ of KIC~8462852~A is given as:
\begin{equation}
    P(\Sigma,\theta) = 1 - e^{-\pi\Sigma\theta^2} = 2.4\times10^{-8}
\end{equation}
}


\revision{Second, we factored in the known demographics of binary companions (i.e., the frequency, mass ratio distribution, and semimajor axis distribution) by modifying the method of \citealt[][hereafter D16]{deacon_pan-starrs_2016}, Appendix A for distinguishing likely binary systems from chance alignments of field stars.  \citetalias{deacon_pan-starrs_2016} Eqn A2 gives the probability of a pair being a true binary pair rather than a coincident pair of field stars as:
\begin{equation}
    P = \frac{\phi_c}{\phi_c+\phi_f}
\end{equation}
where $\phi_c$ and $\phi_f$ are densities for companion and field stars respectively.  While \citetalias{deacon_pan-starrs_2016} considers the full range of binary population, here we are only concerned with the binary fraction that falls within the relevant parameter space, and so we modified \citetalias{deacon_pan-starrs_2016} Eqn A3 to:
\begin{equation}
    \phi_c = f_{bin} \times \frac{1}{A} \times \frac{1}{\Delta m} \times \big[\frac{ e^{ -\Delta\mu^2 / 2\sigma^2_{\mu}}} {2\pi\sigma^2_\mu}\big] \times \big[\frac{ e^{-\Delta \pi^2 / 2\sigma^2_{\pi}}} {\sqrt{2\pi}\sigma_{\pi}}\big]
\end{equation}
where $\Delta\mu$ is relative proper motion in RA/DEC, $\Delta \pi$ is the parallax difference, $\Delta m$ is the size of the magnitude bin used for potential similar companions, $A$ is the total area of the separation bin used, and $f_{bin}$ is the binary companion fraction in those bins.  Using the binary demographics of \citealt[][]{raghavan_survey_2010}, we determined the binary fraction to be $f_{bin} = 0.01$ in a bin of $\Delta \rho = \pm 0.5$ dex of log$_{10}$ projected separation centered on the value for KIC~8462852~B, and $q = \pm 0.05$ in mass ratio between KIC~8462852~AB. We then estimated the corresponding apparent magnitude range ($\delta m_G = \pm 0.6$ mag) using the relations of \citealt[][]{PecautMamajek2016}\footnote{Version 2019.3.22, accessed on 2021-01-06, from \url{http://www.pas.rochester.edu/~emamajek/EEM_dwarf_UBVIJHK_colors_Teff.txt}}.  For $\phi_f$, we performed a {\it Gaia} EDR3 query for all objects within a radius from 50,000 AU (to exclude potential companions) to 30$\degree$ within the same magnitude bin, $+/- 0.5$ mas yr$^{-1}$ proper motion in RA and DEC, and $+/- 0.25$ mas in parallax, returning 132 objects.  We computed the field density as:
\begin{equation}
    \phi_f = \frac{objects}{A \times \Delta m \times (\Delta \mu)^2 \times \Delta \pi}
\end{equation}
We determined a density ratio of
\begin{equation*}
    P = \frac{\phi_c}{\phi_c+\phi_f} = 0.999972
\end{equation*}
}

\revision{Given the extremely high density ratio for a binary companion compared to a field star, and the extremely small probability of chance alignment, we conclude that KIC~8462852~AB is a binary system.
}

\revision{\subsubsection{Test for orbital motion}\label{sec:orbit}}
\revision{Since KIC~8462852~B is a bound companion}, we assume that it follows a Keplerian orbit around the center of mass of the system. 
An object on a circular, face-on orbit at the current 880~AU separation and total system mass of 1.9 M$_\odot$ would have a  velocity $v_{\rm{circ}}$~=~1.4~km~s$^{-1}$, and period P~=~18600~yrs. Our time baseline and measurement precision is therefore marginally capable of measuring linear orbital motion, but is very unlikely to yield a measure of acceleration. Astrometric monitoring alone is unlikely to yield a fit with well-constrained posteriors on the orbit elements, though a refined measure of the linear motion might offer meaningful limits on the joint values of some elements.

To verify this conclusion, we performed a fit to our astrometry for Keplerian orbital elements using our custom implementation of the Orbits for the Impatient (OFTI) rejection sampling algorithm \citep{blunt_orbits_2017}.  OFTI is well-suited to fit poorly constrained astrometric orbits with only small orbit fractions observed for which a Markov Chain Monte Carlo algorithm might not converge \citep{blunt_orbitize_2020}.  OFTI is described in detail in \citealt{blunt_orbits_2017}, and our implementation in \citealt{Pearce2019GSC6214-210}.  In brief, OFTI generates a random set of orbital elements from prior probability distributions, scales the semi-major axis and longitude of periastron to match observations, computes a $\chi^2$ probability for each scaled orbit, and accepts an orbit if the probability of the orbit exceeds a randomly chosen uniform number on the interval [0,1]. We used a total system mass of 1.9 $\pm$ 0.2 \Msun, based on the the mass estimates on Table \ref{table:stellar parameters}. 
We ran our OFTI fitter until 100,000 orbits were accepted.

\begin{figure*}
\centering
\includegraphics[width=0.99\textwidth]{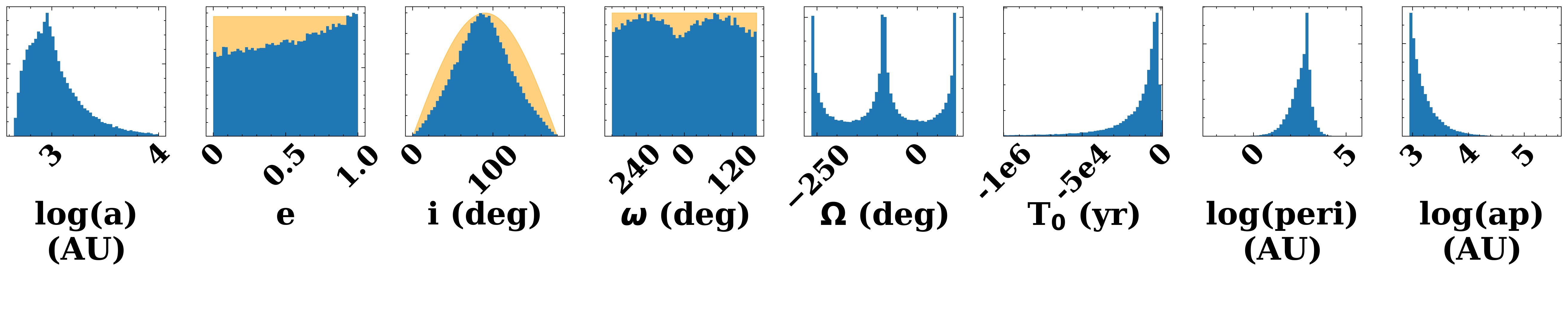}
\caption{\small{ Orbital parameter posterior distributions for KIC~8462852~B.  Posterior distributions on eccentricity, inclination, and argument of periastron are similar to priors \, shown in orange. Semi-major axis and longitude of nodes have no prior due to the scale-and-rotate process of OFTI (see \citealt{blunt_orbits_2017}), while T$_0$, periastron, and apastron are derived from orbital parameters.  
}}
\label{fig:hists_B}
\end{figure*}

Figure \ref{fig:hists_B} displays the posterior distributions for orbital elements, and periastron and apastron distances in our 100,000 orbit sample.  The posterior distributions of orbital elements are similar to priors and do not meaningfully constrain the orbit of KIC~8462852~B relative to A.  We also note that the data did not rule out high eccentricity orbits with extreme values of apastron and periastron, \revision{however the apparent prominence of high eccentricity values is likely a reflection of the uniform eccentricity prior.} \revision{We do not interpret this fit to reveal anything physically meaningful about the orbital elements due to the poor constraint}.

\subsection{ KIC~8462852 cc1}

\begin{figure*}
\centering
\includegraphics[width=0.48\textwidth]{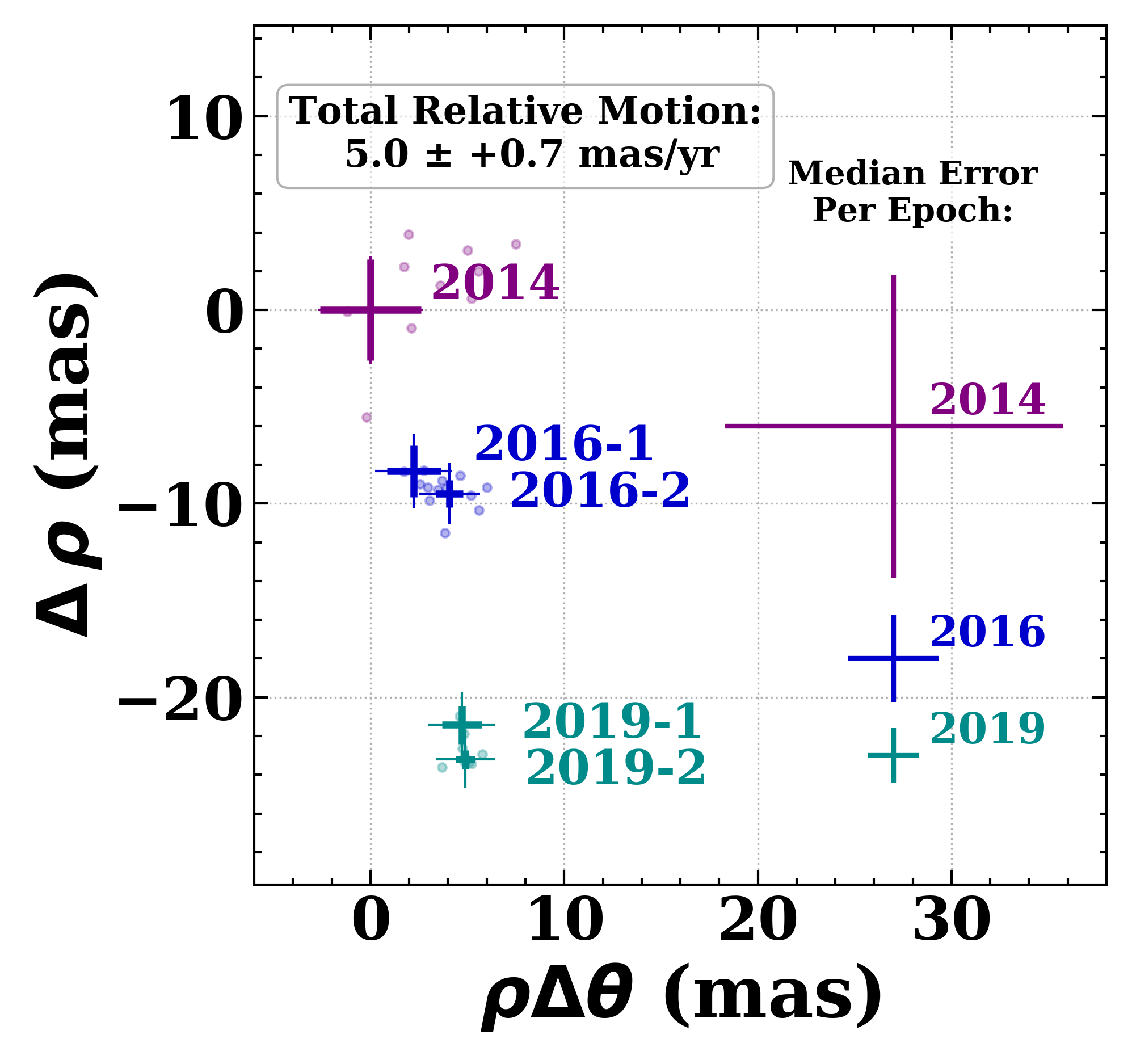}
\includegraphics[width=0.45\textwidth]{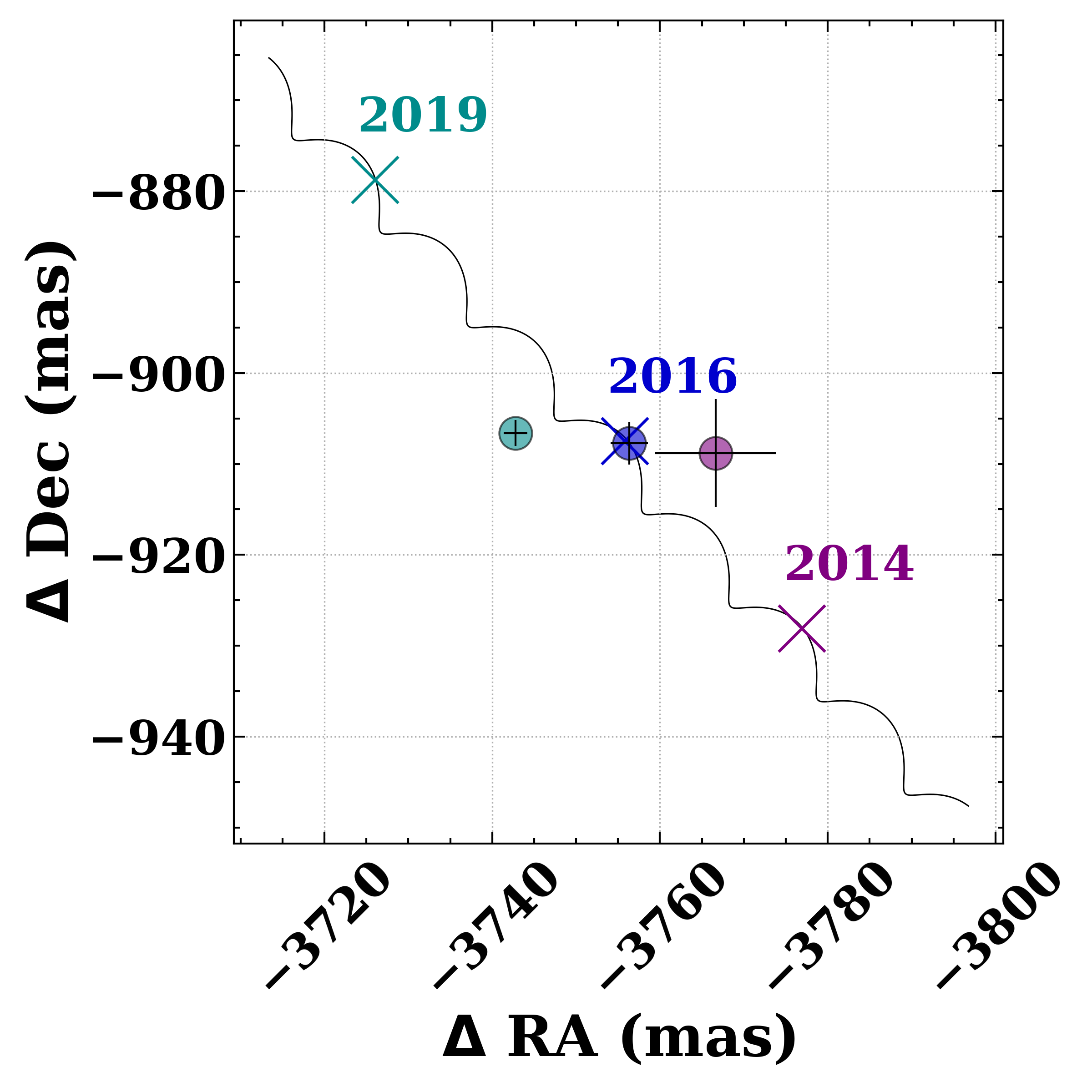}
\caption{\small{Left: Relative astrometry for KIC~8462852 cc1 in individual images (circles) and mean values in each epoch (crosses). Thick crosses show statistical uncertainty, thin crosses show systematic uncertainty.  Crosses to the left display the median error in individual image measurements.  We measure a total relative velocity of $\mu = 5.0 \pm 0.7$ mas yr$^{-1}$.  Epochs 2016 and 2019 dither positions are reported separately.  Right: Observed position of KIC~8462852 cc1 (circles) with expected motion if it were a background star (black track, crosses indicate expected position at observation times).  The relative motion of KIC~8462852 cc1 is not consistent with being a bound companion. It is likely a star with similar space velocity and chance alignment.}}
\label{fig:astr_cc1}
\end{figure*}

Figure \ref{fig:astr_cc1} (left) displays the astrometry results for KIC~8462852 cc1.  We measure a relative proper motion in the plane of the sky of $\Delta\mu = 5.0 \pm 0.7$ mas yr$^{-1}$ ($\Delta\mu = 10.4 \pm 1.5$ km s$^{-1}$), which exceeds the circular velocity at that separation by 6$\sigma$ ($v_{\rm{circ}}$~$\approx$~1~km~s$^{-1}$), and is not consistent with being a bound companion.  Figure \ref{fig:astr_cc1} (right) shows that its relative motion is not consistent with being bound, nor with a non-moving background star.  In testing for co-movement, our measurements fail to reject the null hypothesis that the object is a non-moving background star, yet neither is it consistent with being co-moving ($\chi^2_{\rm{non-moving}} = 500$, $\chi^2_{\rm{comoving}} = 250$).  

Additionally, we performed a search of all objects in \textsl{Gaia} DR2 within 0.5$^{\circ}$ of KIC~8462852 A, displayed in Figure \ref{fig:pm_comparison}, with our three candidate companions.  While the proper motion of KIC~8462852~AB is distinct from the majority, cc1 is similar to the other objects with chance alignment.
We conclude it is most likely a star with similar space velocity and a chance alignment.

\subsection{ KIC~8462852 cc2}

\begin{figure}
\centering
\includegraphics[width=0.46\textwidth]{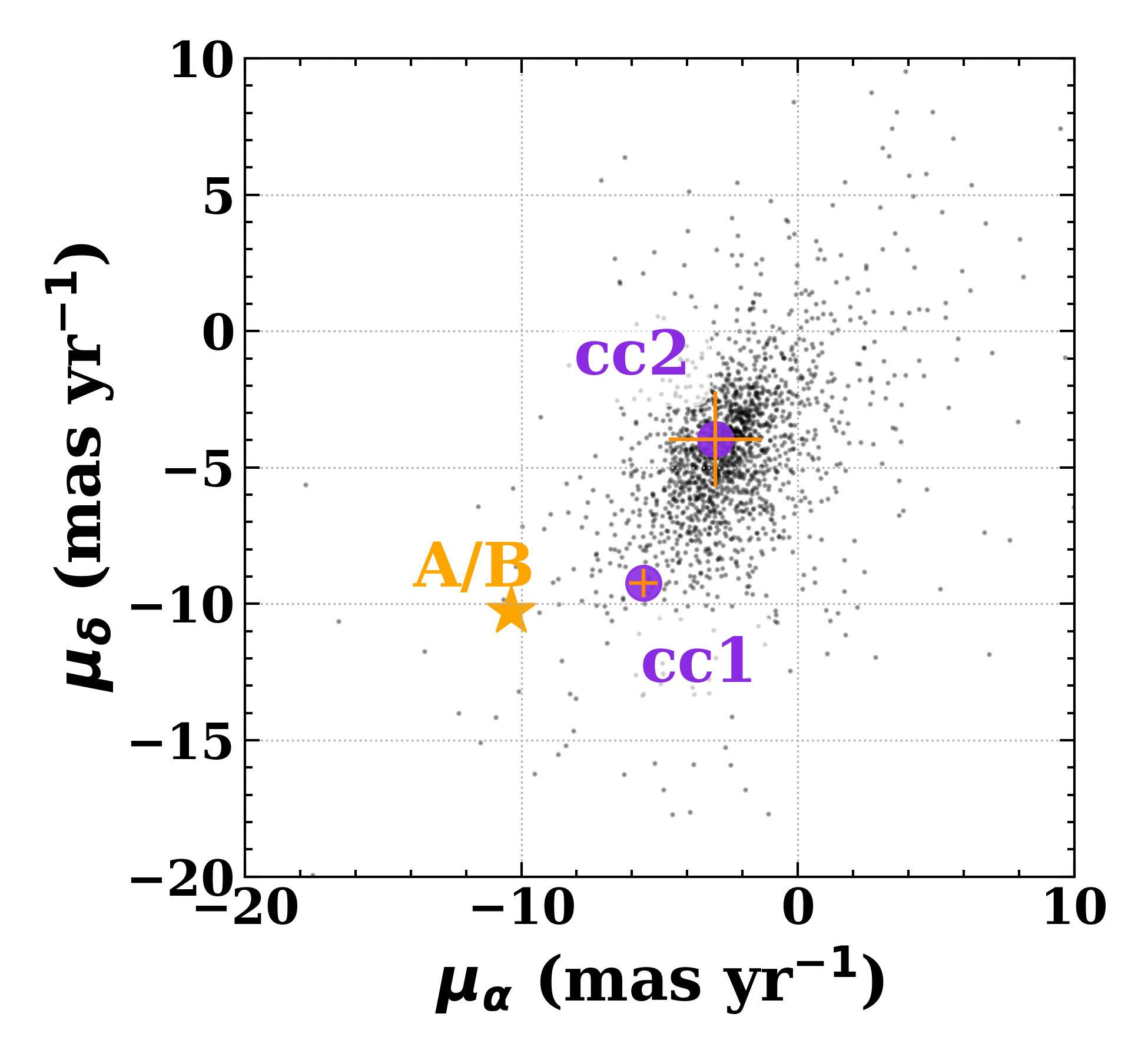}
\caption{\small{The absolute proper motions of KIC~8462852 A, B, cc1, and cc2, and objects in \textsl{Gaia} DR2 in a 0.5$^{\circ}$ cone search around the position of KIC~8462852 A.  The proper motion for cc1 and cc2 are consistent with the motion of nearby stars with chance alignment.  
}}
\label{fig:pm_comparison}
\end{figure}

\begin{figure*}
\centering
\includegraphics[width=0.57\textwidth]{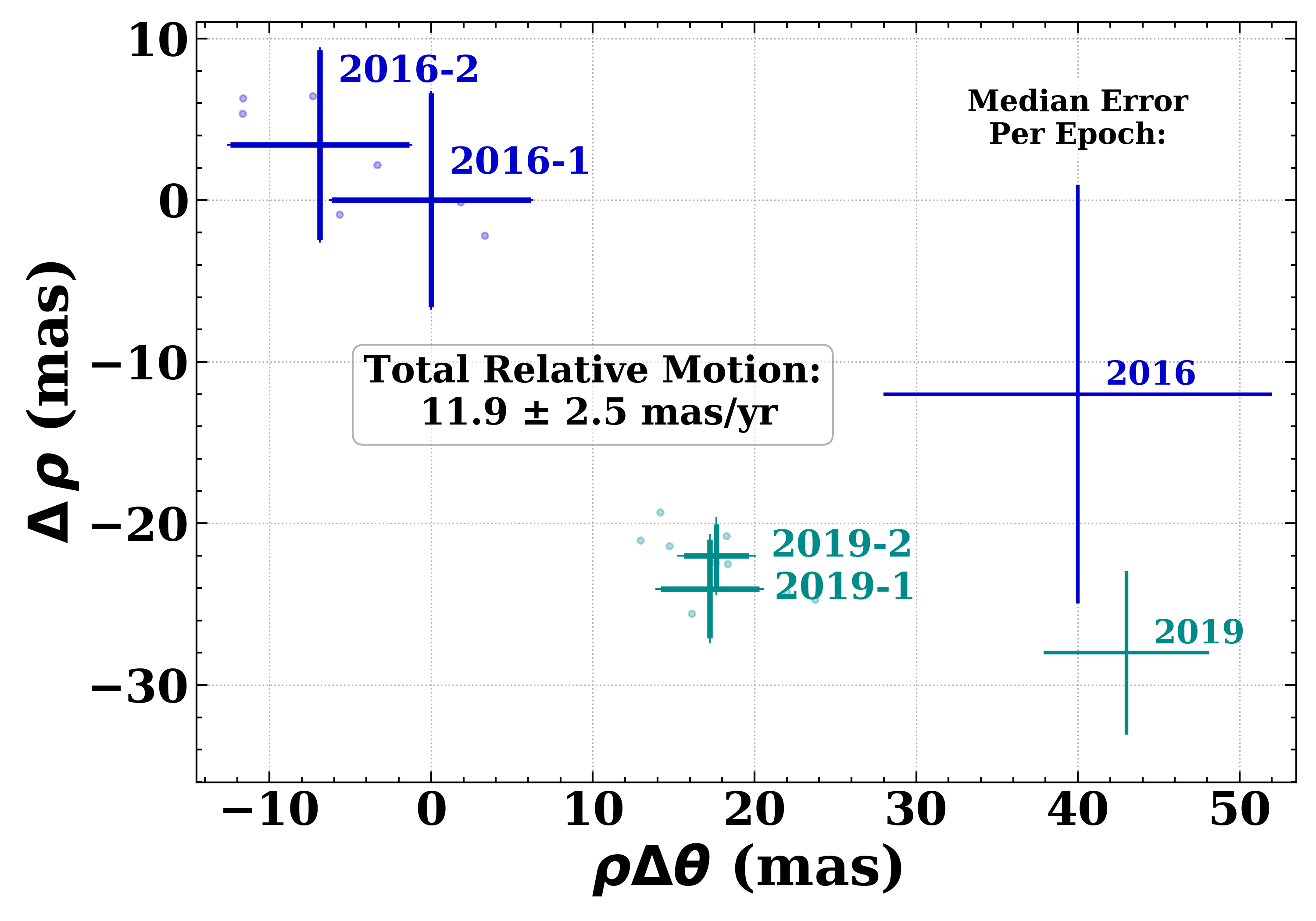}
\includegraphics[width=0.4\textwidth]{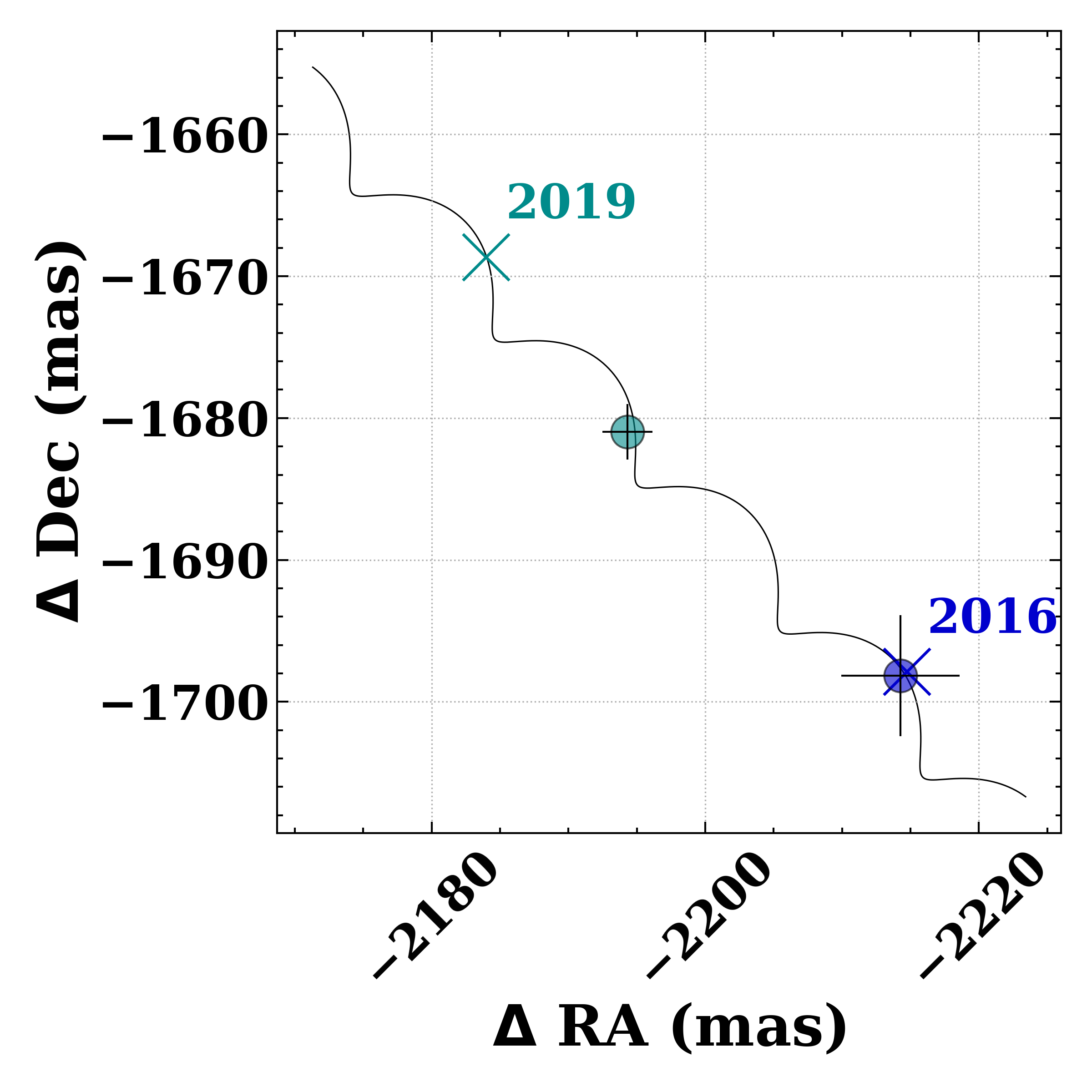}
\caption{\small{Left: Relative astrometry for KIC~8462852 cc2 in individual images (circles) and mean values in each epoch (crosses). Statistical uncertainty dominates systematic uncertainty for this object.  Crosses to the left display the median error in individual image measurements.  We measure a total relative velocity of $\mu = 11.9 \pm 2.5$ mas yr$^{-1}$.  Right: Observed position of KIC~8462852 cc2 (circles) with expected motion if it were a background star (black track, crosses indicate expected position at observation times).  The relative motion of KIC~8462852 cc2 is consistent with an unbound field object.}}
\label{fig:astr_cc2}
\end{figure*}

Figure \ref{fig:astr_cc2} (left) displays the astrometry results for KIC~8462852 cc2.  We measure a relative proper motion in the plane of the sky of $\Delta\mu = 11.9 \pm 2.5$ mas yr$^{-1}$ ($\Delta\mu = 25.2 \pm 5.2$ km s$^{-1}$), which exceeds the circular velocity at that separation by $>5\sigma$ ($v_{\rm{circ}}$~$\approx$~1.2~km~s$^{-1}$), and is not consistent with being a bound companion.  Figure \ref{fig:astr_cc2} (right) shows that its relative motion is more consistent with zero proper motion than with zero relative motion.  Our measurements more strongly support that it is a background object than co-moving, but are not consistent with being completely non-moving ($\chi^2_{\rm{non-moving}} = 150$, $\chi^2_{\rm{comoving}} = 1670$).
The motion of cc2 is consistent with the majority of nearby objects, shown in Figure \ref{fig:pm_comparison}. The \textsl{Gaia} DR2 objects have a mean proper motion of 4.8 mas yr$^{-1}$, within 1$\sigma$ of cc2's absolute proper motion of 5.1$\pm$2.4 mas yr$^{-1}$.  We conclude that cc2 is an unassociated, distant object that is aligned by chance.

\section{Discussion}\label{sec:discussion}

Given the current 880AU projected separation of KIC~8462852~B, it is unlikely to be currently directly influencing the light curve of KIC~8462852~A. However, the binary companion may influence the long-term evolution of the system.  Simulations of wide binary systems have found the Milky Way galactic potential and stellar flybys have significant effect on these systems.  \citet{Kaib2013_galactic_perturbations} found that wide binaries pass through phases of very low pericenter distances ($\sim$100 AU) several times over the course of 10 Gyr due to galactic tides and passing stars, which propagate to disrupting eccentricities of planets and small bodies around one of the stars. 
\citet{Correa-Otto2017_new_insight} found a similar result using an analytic model of the galactic potential. \citet{Correa-Otto2017_galactic_perturbations} built on this numerically to find that a common configuration after 10 Gyr is high eccentricity with semi-major axis from 2000-5000 AU, regardless of the initial orbital configuration.
\citet{Bazs__2020_shadows} found that secular resonances can arise in the habitable zone of stars in binaries even wider than 1000AU if there is also a giant planet present, and the effect of the galactic tide and stellar flybys can push the habitable zone into a high-eccentricity or chaotic state.  There are many combinations of parameter space for which there are not stable orbits, and bodies are disrupted due to secular and mean motion resonances, with significant areas of chaotic orbits.  If the secondary's pericenter is on the move as suggested by \citet{Kaib2013_galactic_perturbations}, disruption of objects in formerly stable orbits is possible as the location of resonances change, as has also been investigated by \citet{Bancelin2019_passing-stars-perturbed-binaries} and \citet{Zakamska2004_excitiation}.

The prospect of long-term orbital evolution for wide binaries suggests that KIC~8462852~B may play a role in the evolution and disruption of stable orbits of objects around the primary. Given the estimated age of this system, this binary may have already undergone at least one phase of close pericenter passage, commonly occurring around 1 Gyr \citep{Kaib2013_galactic_perturbations}.  Our astrometry does not prohibit high-eccentricity/low-pericenter orbits for the binary currently.  
A current or recent low-pericenter phase could excite the eccentricities of planets around A, and disrupt small bodies in the system.  Further astrometric monitoring will continue to improve the picture of the potential influence of the binary on the system.

\revision{We have shown that the pair are not a chance alignment of unassociated stars, but we cannot yet confirm that they follow a bound Keplerian orbit.  It remains possible KIC~8462852~B is a recently ejected star, which which could explain the apparent chaos in the system. However testing this would require a longer astrometric time baseline and a firm measurement of whether high-eccentricity orbits are preferred or disallowed.
It is also possible that the stars are separate members of the same moving group that have remained in close proximity over time.  However, Figure \ref{fig:pm_comparison} and our analysis in Section \ref{sec: binarity} show that there are very few objects with similar proper motions in the vicinity.  In {\it Gaia} EDR3 the closest object with similar proper motion and parallax beyond KIC~8462852~B is two orders of magnitude further in separation, and in Section \ref{sec: binarity} we found a density of $\rho = 9$ objects (mas yr$^{-1}$)$^{-2}$ $\cdot$ mas$^{-1}$ $\cdot$ degree$^{-2}$ with similar parameters to KIC~8462852~B.  Given the 1.2 Gyr age of the system, the population that it formed in should be well dispersed, which is born out by {\it Gaia} astrometry.  Given the small projected separation ($\rho = 880$ AU), we conclude that this explanation is unlikely.   }

\section{Conclusion}\label{sec:conclusion}
We have shown that KIC~8462852~B is a common proper motion pair, and extremely likely to be a gravitationally bound companion to KIC~846285~A using astrometry from Keck/NIRC2 imaging spanning five years.  The relative velocity is consistent with zero during this period.  The time baseline was not long enough to provide meaningful constraints to the pair's orbit, however our analysis shows that they do represent a wide binary pair \revision{rather than chance alignment of field stars}.  We have also shown that two other objects in imaging data are not associated.  Although it has not been thought to be a likely explanation for Boyajian's Star A's light curve, it is a potential source of instability in the long-term evolution of the system, and could excite chaotic orbits of objects in the system.  Efforts to explain KIC~8462852~A's dimming events should be informed by the existence of a wide stellar binary companion to the system.

\acknowledgments

\revision{The authors wish to thank the referee for helpful and constructive comments on this manuscript.}

Much of the code used in this analysis is publicly available at \url{https://github.com/logan-pearce/}

L.A.P.~acknowledges research support from the NSF Graduate Research Fellowship.  This material is based upon work supported by the National Science Foundation Graduate Research Fellowship Program under Grant No. DGE-1746060. 

T.J.D.~acknowledges research support from Gemini Observatory, which is operated by the Association of Universities for Research in Astronomy, Inc., on behalf of the international Gemini partnership of Argentina, Brazil, Canada, Chile, the Republic of Korea, and the United States of America.

D.H. acknowledges support from the Alfred P. Sloan Foundation and the National Aeronautics and Space Administration (80NSSC19K0597).

This work was supported by a NASA Keck PI Data Award, administered by the NASA Exoplanet Science Institute. The data presented herein were obtained at the W. M. Keck Observatory, which is operated as a scientific partnership among the California Institute of Technology, the University of California and the National Aeronautics and Space Administration. The Observatory was made possible by the generous financial support of the W. M. Keck Foundation.  This research has also made use of the Keck Observatory Archive (KOA), which is operated by the W. M. Keck Observatory and the NASA Exoplanet Science Institute (NExScI), under contract with the National Aeronautics and Space Administration.  

The authors wish to recognize and acknowledge the very significant cultural role and reverence that the summit of Maunakea has always had within the indigenous Hawaiian community. We are most fortunate to have the opportunity to conduct observations from this mountain.

The authors acknowledge the Texas Advanced Computing Center (TACC) at The University of Texas at Austin for providing high performance computing resources that have contributed to the research results reported within this paper. URL: http://www.tacc.utexas.edu.

This work has made use of data from the European Space Agency (ESA) mission
{\it Gaia} (\url{https://www.cosmos.esa.int/gaia}), processed by the {\it Gaia}
Data Processing and Analysis Consortium (DPAC,
\url{https://www.cosmos.esa.int/web/gaia/dpac/consortium}). Funding for the DPAC
has been provided by national institutions, in particular the institutions
participating in the {\it Gaia} Multilateral Agreement.

\software{\revision{Numpy \citep{Harris2020Numpy}}, Astropy \citep{astropy:2018}, Matplotlib \citep{Hunter:2007Matplotlib}}

\facilities{Keck:II (ESI), TACC}


\bibliography{refs,references}

\begin{thebibliography}{}
\expandafter\ifx\csname natexlab\endcsname\relax\def\natexlab#1{#1}\fi
\providecommand{\url}[1]{\href{#1}{#1}}

\bibitem[{Bailer-Jones {et~al.}(2018)Bailer-Jones, Rybizki, Fouesneau,
  Mantelet, \& Andrae}]{bailer-jones_estimating_2018}
Bailer-Jones, C. A.~L., Rybizki, J., Fouesneau, M., Mantelet, G., \& Andrae, R.
  2018, The Astronomical Journal, 156, 58.
\newblock \url{http://adsabs.harvard.edu/abs/2018AJ....156...58B}

\bibitem[{{Bancelin} {et~al.}(2019){Bancelin}, {Nordlander}, {Pilat-Lohinger},
  \& {Loibnegger}}]{Bancelin2019_passing-stars-perturbed-binaries}
{Bancelin}, D., {Nordlander}, T., {Pilat-Lohinger}, E., \& {Loibnegger}, B.
  2019, \mnras, 486, 4773

\bibitem[{Bazs{\'{o}} \& Pilat-Lohinger(2020)}]{Bazs__2020_shadows}
Bazs{\'{o}}, {\'{A}}., \& Pilat-Lohinger, E. 2020, The Astronomical Journal,
  160, 2.
\newblock \url{https://doi.org/10.3847%2F1538-3881%2Fab9104}

\bibitem[{{Berger} {et~al.}(2020){Berger}, {Huber}, {van Saders}, {Gaidos},
  {Tayar}, \& {Kraus}}]{berger20}
{Berger}, T.~A., {Huber}, D., {van Saders}, J.~L., {et~al.} 2020, \aj, 159, 280

\bibitem[{Blunt {et~al.}(2017)Blunt, Nielsen, De~Rosa, Konopacky, Ryan, Wang,
  Pueyo, Rameau, Marois, Marchis, Macintosh, Graham, Duchêne, \&
  Schneider}]{blunt_orbits_2017}
Blunt, S., Nielsen, E.~L., De~Rosa, R.~J., {et~al.} 2017, The Astronomical
  Journal, 153, 229.
\newblock \url{http://adsabs.harvard.edu/abs/2017AJ....153..229B}

\bibitem[{Blunt {et~al.}(2020)Blunt, Wang, Angelo, Ngo, Cody, De~Rosa, Graham,
  Hirsch, Nagpal, Nielsen, Pearce, Rice, \& Tejada}]{blunt_orbitize_2020}
Blunt, S., Wang, J.~J., Angelo, I., {et~al.} 2020, The Astronomical Journal,
  159, 89.
\newblock \url{http://adsabs.harvard.edu/abs/2020AJ....159...89B}

\bibitem[{Bodman \& Quillen(2016)}]{bodman_kic_2016}
Bodman, E. H.~L., \& Quillen, A. 2016, The Astrophysical Journal Letters, 819,
  L34.
\newblock \url{http://adsabs.harvard.edu/abs/2016ApJ...819L..34B}

\bibitem[{Boyajian {et~al.}(2016)Boyajian, LaCourse, Rappaport, Fabrycky,
  Fischer, Gandolfi, Kennedy, Korhonen, Liu, Moor, Olah, Vida, Wyatt, Best,
  Brewer, Ciesla, Csák, Deeg, Dupuy, Handler, Heng, Howell, Ishikawa, Kovács,
  Kozakis, Kriskovics, Lehtinen, Lintott, Lynn, Nespral, Nikbakhsh, Schawinski,
  Schmitt, Smith, Szabo, Szabo, Viuho, Wang, Weiksnar, Bosch, Connors, Goodman,
  Green, Hoekstra, Jebson, Jek, Omohundro, Schwengeler, \&
  Szewczyk}]{boyajian_planet_2016}
Boyajian, T.~S., LaCourse, D.~M., Rappaport, S.~A., {et~al.} 2016, Monthly
  Notices of the Royal Astronomical Society, 457, 3988.
\newblock \url{http://adsabs.harvard.edu/abs/2016MNRAS.457.3988B}

\bibitem[{Boyajian {et~al.}(2018)Boyajian, Alonso, Ammerman, Armstrong,
  Asensio~Ramos, Barkaoui, Beatty, Benkhaldoun, Benni, Bentley, Berdyugin,
  Berdyugina, Bergeron, Bieryla, Blain, Capetillo~Blanco, Bodman, Boucher,
  Bradley, Brincat, Brink, Briol, Brown, Budaj, Burdanov, Cale, Aznar~Carbo,
  Castillo~García, Clark, Clayton, Clem, Coker, Cook, Copperwheat, Curtis,
  Cutri, Cseh, Cynamon, Daniels, Davenport, Deeg, De~Lorenzo, de~Jaeger,
  Desrosiers, Dolan, Dowhos, Dubois, Durkee, Dvorak, Easley, Edwards, Ellis,
  Erdelyi, Ertel, Farfán, Farihi, Filippenko, Foxell, Gandolfi, Garcia,
  Giddens, Gillon, González-Carballo, González-Fernández,
  González~Hernández, Graham, Greene, Gregorio, Hallakoun, Hanyecz, Harp,
  Henry, Herrero, Hildbold, Hinzel, Holgado, Ignácz, Ilyin, Ivanov, Jehin,
  Jermak, Johnston, Kafka, Kalup, Kardasis, Kaspi, Kennedy, Kiefer, Kielty,
  Kessler, Kiiskinen, Killestein, King, Kollar, Korhonen, Kotnik,
  Könyves-Tóth, Kriskovics, Krumm, Krushinsky, Kundra, Lachapelle, LaCourse,
  Lake, Lam, Lamb, Lane, Lau, Lewin, Lintott, Lisse, Logie, Longeard,
  Lopez~Villanueva, Whit~Ludington, Mainzer, Malo, Maloney, Mann, Mantero,
  Marengo, Marchant, Martínez~González, Masiero, Mauerhan, McCormac, McNeely,
  Meng, Miller, Molnar, Morales, Morris, Muterspaugh, Nespral, Nugent, Nugent,
  Odasso, O'Keeffe, Oksanen, O'Meara, Ordasi, Osborn, Ott, Parks,
  Rodriguez~Perez, Petriew, Pickard, Pál, Plavchan, Pollacco, Pozo~Nuñez,
  Pozuelos, Rau, Redfield, Relles, Ribas, Richards, Saario, Safron, Sallai,
  Sárneczky, Schaefer, Schumer, Schwartzendruber, Siegel, Siemion, Simmons,
  Simon, Simón-Díaz, Sitko, Socas-Navarro, Sódor, Starkey, Steele, Stone,
  Strassmeier, Street, Sullivan, Suomela, Swift, Szabó, Szabó, Szakáts,
  Szalai, Tanner, Toledo-Padrón, Tordai, Triaud, Turner, Ulowetz, Urbanik,
  Vanaverbeke, Vanderburg, Vida, Vietje, Vinkó, von Braun, Waagen, Walsh,
  Watson, Weir, Wenzel, Westendorp~Plaza, Williamson, Wright, Wyatt, Zheng, \&
  Zsidi}]{boyajian_first_2018}
Boyajian, T.~S., Alonso, R., Ammerman, A., {et~al.} 2018, The Astrophysical
  Journal Letters, 853, L8.
\newblock \url{http://adsabs.harvard.edu/abs/2018ApJ...853L...8B}

\bibitem[{{Choi} {et~al.}(2016){Choi}, {Dotter}, {Conroy}, {Cantiello},
  {Paxton}, \& {Johnson}}]{choi16}
{Choi}, J., {Dotter}, A., {Conroy}, C., {et~al.} 2016, \apj, 823, 102

\bibitem[{Clemens {et~al.}(2018)Clemens, Maheshwari, Jagani, Montgomery,
  El~Batal, Ellis, \& Wright}]{clemens_proper_2018}
Clemens, D.~P., Maheshwari, K., Jagani, R., {et~al.} 2018, The Astrophysical
  Journal Letters, 856, L8.
\newblock \url{http://adsabs.harvard.edu/abs/2018ApJ...856L...8C}

\bibitem[{{Correa-Otto} {et~al.}(2017){Correa-Otto}, {Calandra}, \&
  {Gil-Hutton}}]{Correa-Otto2017_new_insight}
{Correa-Otto}, J.~A., {Calandra}, M.~F., \& {Gil-Hutton}, R.~A. 2017, \aap,
  600, A59

\bibitem[{{Correa-Otto} \&
  {Gil-Hutton}(2017)}]{Correa-Otto2017_galactic_perturbations}
{Correa-Otto}, J.~A., \& {Gil-Hutton}, R.~A. 2017, \aap, 608, A116

\bibitem[{{Correia} {et~al.}(2006){Correia}, {Zinnecker}, {Ratzka}, \&
  {Sterzik}}]{Correia2006VisualPMSBinaries}
{Correia}, S., {Zinnecker}, H., {Ratzka}, T., \& {Sterzik}, M.~F. 2006, \aap,
  459, 909

\bibitem[{Deacon {et~al.}(2016)Deacon, Kraus, Mann, Magnier, Chambers,
  Wainscoat, Tonry, Kaiser, Waters, Flewelling, Hodapp, \&
  Burgett}]{deacon_pan-starrs_2016}
Deacon, N.~R., Kraus, A.~L., Mann, A.~W., {et~al.} 2016, Monthly Notices of the
  Royal Astronomical Society, 455, 4212.
\newblock \url{http://adsabs.harvard.edu/abs/2016MNRAS.455.4212D}

\bibitem[{Dupuy {et~al.}(2016)Dupuy, Kratter, Kraus, Isaacson, Mann, Ireland,
  Howard, \& Huber}]{dupuy_orbital_2016}
Dupuy, T.~J., Kratter, K.~M., Kraus, A.~L., {et~al.} 2016, The Astrophysical
  Journal, 817, 80.
\newblock \url{http://adsabs.harvard.edu/abs/2016ApJ...817...80D}

\bibitem[{{Gaia Collaboration} {et~al.}(2020){Gaia Collaboration}, {Brown},
  {Vallenari}, {Prusti}, {de Bruijne}, {Babusiaux}, \&
  {Biermann}}]{GaiaEDR3summary}
{Gaia Collaboration}, {Brown}, A.~G.~A., {Vallenari}, A., {et~al.} 2020, arXiv
  e-prints, arXiv:2012.01533

\bibitem[{{Gaia Collaboration} {et~al.}(2016){Gaia Collaboration}, Prusti,
  de~Bruijne, Brown, Vallenari, Babusiaux, Bailer-Jones, Bastian, Biermann,
  Evans, Eyer, Jansen, Jordi, Klioner, Lammers, Lindegren, Luri, Mignard,
  Milligan, Panem, Poinsignon, Pourbaix, Randich, Sarri, Sartoretti, Siddiqui,
  Soubiran, Valette, van Leeuwen, Walton, Aerts, Arenou, Cropper, Drimmel,
  Høg, Katz, Lattanzi, O'Mullane, Grebel, Holland, Huc, Passot, Bramante,
  Cacciari, Castañeda, Chaoul, Cheek, De~Angeli, Fabricius, Guerra,
  Hernández, Jean-Antoine-Piccolo, Masana, Messineo, Mowlavi, Nienartowicz,
  Ordóñez-Blanco, Panuzzo, Portell, Richards, Riello, Seabroke, Tanga,
  Thévenin, Torra, Els, Gracia-Abril, Comoretto, Garcia-Reinaldos, Lock,
  Mercier, Altmann, Andrae, Astraatmadja, Bellas-Velidis, Benson, Berthier,
  Blomme, Busso, Carry, Cellino, Clementini, Cowell, Creevey, Cuypers,
  Davidson, De~Ridder, de~Torres, Delchambre, Dell'Oro, Ducourant, Frémat,
  García-Torres, Gosset, Halbwachs, Hambly, Harrison, Hauser, Hestroffer,
  Hodgkin, Huckle, Hutton, Jasniewicz, Jordan, Kontizas, Korn, Lanzafame,
  Manteiga, Moitinho, Muinonen, Osinde, Pancino, Pauwels, Petit, Recio-Blanco,
  Robin, Sarro, Siopis, Smith, Smith, Sozzetti, Thuillot, van Reeven, Viala,
  Abbas, Abreu~Aramburu, Accart, Aguado, Allan, Allasia, Altavilla, Álvarez,
  Alves, Anderson, Andrei, Anglada~Varela, Antiche, Antoja, Antón, Arcay,
  Atzei, Ayache, Bach, Baker, Balaguer-Núñez, Barache, Barata, Barbier,
  Barblan, Baroni, Barrado~y Navascués, Barros, Barstow, Becciani, Bellazzini,
  Bellei, Bello~García, Belokurov, Bendjoya, Berihuete, Bianchi, Bienaymé,
  Billebaud, Blagorodnova, Blanco-Cuaresma, Boch, Bombrun, Borrachero,
  Bouquillon, Bourda, Bouy, Bragaglia, Breddels, Brouillet, Brüsemeister,
  Bucciarelli, Budnik, Burgess, Burgon, Burlacu, Busonero, Buzzi, Caffau,
  Cambras, Campbell, Cancelliere, Cantat-Gaudin, Carlucci, Carrasco,
  Castellani, Charlot, Charnas, Charvet, Chassat, Chiavassa, Clotet, Cocozza,
  Collins, Collins, Costigan, Crifo, Cross, Crosta, Crowley, Dafonte, Damerdji,
  Dapergolas, David, David, De~Cat, de~Felice, de~Laverny, De~Luise, De~March,
  de~Martino, de~Souza, Debosscher, del Pozo, Delbo, Delgado, Delgado,
  di~Marco, Di~Matteo, Diakite, Distefano, Dolding, Dos~Anjos, Drazinos,
  Durán, Dzigan, Ecale, Edvardsson, Enke, Erdmann, Escolar, Espina, Evans,
  Eynard~Bontemps, Fabre, Fabrizio, Faigler, Falcão, Farràs~Casas, Faye,
  Federici, Fedorets, Fernández-Hernández, Fernique, Fienga, Figueras,
  Filippi, Findeisen, Fonti, Fouesneau, Fraile, Fraser, Fuchs, Furnell, Gai,
  Galleti, Galluccio, Garabato, García-Sedano, Garé, Garofalo, Garralda,
  Gavras, Gerssen, Geyer, Gilmore, Girona, Giuffrida, Gomes, González-Marcos,
  González-Núñez, González-Vidal, Granvik, Guerrier, Guillout, Guiraud,
  Gúrpide, Gutiérrez-Sánchez, Guy, Haigron, Hatzidimitriou, Haywood, Heiter,
  Helmi, Hobbs, Hofmann, Holl, Holland, Hunt, Hypki, Icardi, Irwin, Jevardat~de
  Fombelle, Jofré, Jonker, Jorissen, Julbe, Karampelas, Kochoska, Kohley,
  Kolenberg, Kontizas, Koposov, Kordopatis, Koubsky, Kowalczyk, Krone-Martins,
  Kudryashova, Kull, Bachchan, Lacoste-Seris, Lanza, Lavigne,
  Le~Poncin-Lafitte, Lebreton, Lebzelter, Leccia, Leclerc, Lecoeur-Taibi,
  Lemaitre, Lenhardt, Leroux, Liao, Licata, Lindstrøm, Lister, Livanou, Lobel,
  Löffler, López, Lopez-Lozano, Lorenz, Loureiro, MacDonald,
  Magalhães~Fernandes, Managau, Mann, Mantelet, Marchal, Marchant, Marconi,
  Marie, Marinoni, Marrese, Marschalkó, Marshall, Martín-Fleitas, Martino,
  Mary, Matijevič, Mazeh, McMillan, Messina, Mestre, Michalik, Millar,
  Miranda, Molina, Molinaro, Molinaro, Molnár, Moniez, Montegriffo, Monteiro,
  Mor, Mora, Morbidelli, Morel, Morgenthaler, Morley, Morris, Mulone, Muraveva,
  Musella, Narbonne, Nelemans, Nicastro, Noval, Ordénovic, Ordieres-Meré,
  Osborne, Pagani, Pagano, Pailler, Palacin, Palaversa, Parsons, Paulsen,
  Pecoraro, Pedrosa, Pentikäinen, Pereira, Pichon, Piersimoni, Pineau, Plachy,
  Plum, Poujoulet, Prša, Pulone, Ragaini, Rago, Rambaux, Ramos-Lerate,
  Ranalli, Rauw, Read, Regibo, Renk, Reylé, Ribeiro, Rimoldini, Ripepi, Riva,
  Rixon, Roelens, Romero-Gómez, Rowell, Royer, Rudolph, Ruiz-Dern, Sadowski,
  Sagristà~Sellés, Sahlmann, Salgado, Salguero, Sarasso, Savietto, Schnorhk,
  Schultheis, Sciacca, Segol, Segovia, Segransan, Serpell, Shih, Smareglia,
  Smart, Smith, Solano, Solitro, Sordo, Soria~Nieto, Souchay, Spagna, Spoto,
  Stampa, Steele, Steidelmüller, Stephenson, Stoev, Suess, Süveges, Surdej,
  Szabados, Szegedi-Elek, Tapiador, Taris, Tauran, Taylor, Teixeira, Terrett,
  Tingley, Trager, Turon, Ulla, Utrilla, Valentini, van Elteren, Van~Hemelryck,
  van Leeuwen, Varadi, Vecchiato, Veljanoski, Via, Vicente, Vogt, Voss,
  Votruba, Voutsinas, Walmsley, Weiler, Weingrill, Werner, Wevers, Whitehead,
  Wyrzykowski, Yoldas, Žerjal, Zucker, Zurbach, Zwitter, Alecu, Allen,
  Allende~Prieto, Amorim, Anglada-Escudé, Arsenijevic, Azaz, Balm, Beck,
  Bernstein, Bigot, Bijaoui, Blasco, Bonfigli, Bono, Boudreault, Bressan,
  Brown, Brunet, Bunclark, Buonanno, Butkevich, Carret, Carrion, Chemin,
  Chéreau, Corcione, Darmigny, de~Boer, de~Teodoro, de~Zeeuw, Delle~Luche,
  Domingues, Dubath, Fodor, Frézouls, Fries, Fustes, Fyfe, Gallardo, Gallegos,
  Gardiol, Gebran, Gomboc, Gómez, Grux, Gueguen, Heyrovsky, Hoar, Iannicola,
  Isasi~Parache, Janotto, Joliet, Jonckheere, Keil, Kim, Klagyivik, Klar,
  Knude, Kochukhov, Kolka, Kos, Kutka, Lainey, LeBouquin, Liu, Loreggia,
  Makarov, Marseille, Martayan, Martinez-Rubi, Massart, Meynadier, Mignot,
  Munari, Nguyen, Nordlander, Ocvirk, O'Flaherty, Olias~Sanz, Ortiz, Osorio,
  Oszkiewicz, Ouzounis, Palmer, Park, Pasquato, Peltzer, Peralta, Péturaud,
  Pieniluoma, Pigozzi, Poels, Prat, Prod'homme, Raison, Rebordao, Risquez,
  Rocca-Volmerange, Rosen, Ruiz-Fuertes, Russo, Sembay, Serraller~Vizcaino,
  Short, Siebert, Silva, Sinachopoulos, Slezak, Soffel, Sosnowska, Straižys,
  ter Linden, Terrell, Theil, Tiede, Troisi, Tsalmantza, Tur, Vaccari, Vachier,
  Valles, Van~Hamme, Veltz, Virtanen, Wallut, Wichmann, Wilkinson, Ziaeepour,
  \& Zschocke}]{gaia_collaboration_gaia_2016}
{Gaia Collaboration}, Prusti, T., de~Bruijne, J. H.~J., {et~al.} 2016,
  Astronomy and Astrophysics, 595, A1.
\newblock \url{http://adsabs.harvard.edu/abs/2016A%26A...595A...1G}

\bibitem[{{Gaia Collaboration} {et~al.}(2018){Gaia Collaboration}, Brown,
  Vallenari, Prusti, de~Bruijne, Babusiaux, Bailer-Jones, Biermann, Evans,
  Eyer, Jansen, Jordi, Klioner, Lammers, Lindegren, Luri, Mignard, Panem,
  Pourbaix, Randich, Sartoretti, Siddiqui, Soubiran, van Leeuwen, Walton,
  Arenou, Bastian, Cropper, Drimmel, Katz, Lattanzi, Bakker, Cacciari,
  Castañeda, Chaoul, Cheek, De~Angeli, Fabricius, Guerra, Holl, Masana,
  Messineo, Mowlavi, Nienartowicz, Panuzzo, Portell, Riello, Seabroke, Tanga,
  Thévenin, Gracia-Abril, Comoretto, Garcia-Reinaldos, Teyssier, Altmann,
  Andrae, Audard, Bellas-Velidis, Benson, Berthier, Blomme, Burgess, Busso,
  Carry, Cellino, Clementini, Clotet, Creevey, Davidson, De~Ridder, Delchambre,
  Dell'Oro, Ducourant, Fernández-Hernández, Fouesneau, Frémat, Galluccio,
  García-Torres, González-Núñez, González-Vidal, Gosset, Guy, Halbwachs,
  Hambly, Harrison, Hernández, Hestroffer, Hodgkin, Hutton, Jasniewicz,
  Jean-Antoine-Piccolo, Jordan, Korn, Krone-Martins, Lanzafame, Lebzelter,
  Löffler, Manteiga, Marrese, Martín-Fleitas, Moitinho, Mora, Muinonen,
  Osinde, Pancino, Pauwels, Petit, Recio-Blanco, Richards, Rimoldini, Robin,
  Sarro, Siopis, Smith, Sozzetti, Süveges, Torra, van Reeven, Abbas,
  Abreu~Aramburu, Accart, Aerts, Altavilla, Álvarez, Alvarez, Alves, Anderson,
  Andrei, Anglada~Varela, Antiche, Antoja, Arcay, Astraatmadja, Bach, Baker,
  Balaguer-Núñez, Balm, Barache, Barata, Barbato, Barblan, Barklem, Barrado,
  Barros, Barstow, Bartholomé~Muñoz, Bassilana, Becciani, Bellazzini,
  Berihuete, Bertone, Bianchi, Bienaymé, Blanco-Cuaresma, Boch, Boeche,
  Bombrun, Borrachero, Bossini, Bouquillon, Bourda, Bragaglia, Bramante,
  Breddels, Bressan, Brouillet, Brüsemeister, Brugaletta, Bucciarelli,
  Burlacu, Busonero, Butkevich, Buzzi, Caffau, Cancelliere, Cannizzaro,
  Cantat-Gaudin, Carballo, Carlucci, Carrasco, Casamiquela, Castellani,
  Castro-Ginard, Charlot, Chemin, Chiavassa, Cocozza, Costigan, Cowell, Crifo,
  Crosta, Crowley, Cuypers, Dafonte, Damerdji, Dapergolas, David, David,
  de~Laverny, De~Luise, De~March, de~Martino, de~Souza, de~Torres, Debosscher,
  del Pozo, Delbo, Delgado, Delgado, Di~Matteo, Diakite, Diener, Distefano,
  Dolding, Drazinos, Durán, Edvardsson, Enke, Eriksson, Esquej,
  Eynard~Bontemps, Fabre, Fabrizio, Faigler, Falcão, Farràs~Casas, Federici,
  Fedorets, Fernique, Figueras, Filippi, Findeisen, Fonti, Fraile, Fraser,
  Frézouls, Gai, Galleti, Garabato, García-Sedano, Garofalo, Garralda, Gavel,
  Gavras, Gerssen, Geyer, Giacobbe, Gilmore, Girona, Giuffrida, Glass, Gomes,
  Granvik, Gueguen, Guerrier, Guiraud, Gutiérrez-Sánchez, Haigron,
  Hatzidimitriou, Hauser, Haywood, Heiter, Helmi, Heu, Hilger, Hobbs, Hofmann,
  Holland, Huckle, Hypki, Icardi, Janßen, Jevardat~de Fombelle, Jonker,
  Juhász, Julbe, Karampelas, Kewley, Klar, Kochoska, Kohley, Kolenberg,
  Kontizas, Kontizas, Koposov, Kordopatis, Kostrzewa-Rutkowska, Koubsky,
  Lambert, Lanza, Lasne, Lavigne, Le~Fustec, Le~Poncin-Lafitte, Lebreton,
  Leccia, Leclerc, Lecoeur-Taibi, Lenhardt, Leroux, Liao, Licata, Lindstrøm,
  Lister, Livanou, Lobel, López, Managau, Mann, Mantelet, Marchal, Marchant,
  Marconi, Marinoni, Marschalkó, Marshall, Martino, Marton, Mary, Massari,
  Matijevič, Mazeh, McMillan, Messina, Michalik, Millar, Molina, Molinaro,
  Molnár, Montegriffo, Mor, Morbidelli, Morel, Morris, Mulone, Muraveva,
  Musella, Nelemans, Nicastro, Noval, O'Mullane, Ordénovic, Ordóñez-Blanco,
  Osborne, Pagani, Pagano, Pailler, Palacin, Palaversa, Panahi, Pawlak,
  Piersimoni, Pineau, Plachy, Plum, Poggio, Poujoulet, Prša, Pulone, Racero,
  Ragaini, Rambaux, Ramos-Lerate, Regibo, Reylé, Riclet, Ripepi, Riva, Rivard,
  Rixon, Roegiers, Roelens, Romero-Gómez, Rowell, Royer, Ruiz-Dern, Sadowski,
  Sagristà~Sellés, Sahlmann, Salgado, Salguero, Sanna, Santana-Ros, Sarasso,
  Savietto, Schultheis, Sciacca, Segol, Segovia, Ségransan, Shih, Siltala,
  Silva, Smart, Smith, Solano, Solitro, Sordo, Soria~Nieto, Souchay, Spagna,
  Spoto, Stampa, Steele, Steidelmüller, Stephenson, Stoev, Suess, Surdej,
  Szabados, Szegedi-Elek, Tapiador, Taris, Tauran, Taylor, Teixeira, Terrett,
  Teyssandier, Thuillot, Titarenko, Torra~Clotet, Turon, Ulla, Utrilla, Uzzi,
  Vaillant, Valentini, Valette, van Elteren, Van~Hemelryck, van Leeuwen,
  Vaschetto, Vecchiato, Veljanoski, Viala, Vicente, Vogt, von Essen, Voss,
  Votruba, Voutsinas, Walmsley, Weiler, Wertz, Wevers, Wyrzykowski, Yoldas,
  Žerjal, Ziaeepour, Zorec, Zschocke, Zucker, Zurbach, \&
  Zwitter}]{gaia_collaboration_gaia_2018}
{Gaia Collaboration}, Brown, A. G.~A., Vallenari, A., {et~al.} 2018, Astronomy
  and Astrophysics, 616, A1.
\newblock \url{http://adsabs.harvard.edu/abs/2018A%26A...616A...1G}

\bibitem[{Harris {et~al.}(2020)Harris, Millman, van~der Walt, Gommers,
  Virtanen, Cournapeau, Wieser, Taylor, Berg, Smith, Kern, Picus, Hoyer, van
  Kerkwijk, Brett, Haldane, del R{'{\i}}o, Wiebe, Peterson,
  G{'{e}}rard-Marchant, Sheppard, Reddy, Weckesser, Abbasi, Gohlke, \&
  Oliphant}]{Harris2020Numpy}
Harris, C.~R., Millman, K.~J., van~der Walt, S.~J., {et~al.} 2020, Nature, 585,
  357.
\newblock \url{https://doi.org/10.1038/s41586-020-2649-2}

\bibitem[{{Huber} {et~al.}(2017){Huber}, {Zinn}, {Bojsen-Hansen},
  {Pinsonneault}, {Sahlholdt}, {Serenelli}, {Silva Aguirre}, {Stassun},
  {Stello}, {Tayar}, {Bastien}, {Bedding}, {Buchhave}, {Chaplin}, {Davies},
  {Garc{\'\i}a}, {Latham}, {Mathur}, {Mosser}, \& {Sharma}}]{huber17}
{Huber}, D., {Zinn}, J., {Bojsen-Hansen}, M., {et~al.} 2017, \apj, 844, 102

\bibitem[{Hunter(2007)}]{Hunter:2007Matplotlib}
Hunter, J.~D. 2007, Computing In Science \& Engineering, 9, 90

\bibitem[{{Kaib} {et~al.}(2013){Kaib}, {Raymond}, \&
  {Duncan}}]{Kaib2013_galactic_perturbations}
{Kaib}, N.~A., {Raymond}, S.~N., \& {Duncan}, M. 2013, \nat, 493, 381

\bibitem[{{Kraus} {et~al.}(2016){Kraus}, {Ireland}, {Huber}, {Mann}, \&
  {Dupuy}}]{stellar}
{Kraus}, A.~L., {Ireland}, M.~J., {Huber}, D., {Mann}, A.~W., \& {Dupuy}, T.~J.
  2016, \aj, 152, 8

\bibitem[{Kraus {et~al.}(2016)Kraus, Ireland, Huber, Mann, \&
  Dupuy}]{kraus_impact_2016}
Kraus, A.~L., Ireland, M.~J., Huber, D., Mann, A.~W., \& Dupuy, T.~J. 2016, The
  Astronomical Journal, 152, 8.
\newblock \url{http://adsabs.harvard.edu/abs/2016AJ....152....8K}

\bibitem[{Lisse {et~al.}(2015)Lisse, Sitko, \& Marengo}]{lisse_irtfspex_2015}
Lisse, C.~M., Sitko, M.~L., \& Marengo, M. 2015, The Astrophysical Journal
  Letters, 815, L27.
\newblock \url{http://adsabs.harvard.edu/abs/2015ApJ...815L..27L}

\bibitem[{{Mann} {et~al.}(2015){Mann}, {Feiden}, {Gaidos}, {Boyajian}, \& {von
  Braun}}]{mann15}
{Mann}, A.~W., {Feiden}, G.~A., {Gaidos}, E., {Boyajian}, T., \& {von Braun},
  K. 2015, \apj, 804, 64

\bibitem[{Mann {et~al.}(2019)Mann, Dupuy, Kraus, Gaidos, Ansdell, Ireland,
  Rizzuto, Hung, Dittmann, Factor, Feiden, Martinez, Ruíz-Rodríguez, \&
  Thao}]{mann_how_2019}
Mann, A.~W., Dupuy, T., Kraus, A.~L., {et~al.} 2019, The Astrophysical Journal,
  871, 63.
\newblock \url{http://adsabs.harvard.edu/abs/2019ApJ...871...63M}

\bibitem[{Marengo {et~al.}(2015)Marengo, Hulsebus, \&
  Willis}]{marengo_kic_2015}
Marengo, M., Hulsebus, A., \& Willis, S. 2015, The Astrophysical Journal
  Letters, 814, L15.
\newblock \url{http://adsabs.harvard.edu/abs/2015ApJ...814L..15M}

\bibitem[{{Mart{\'\i}nez Gonz{\'a}lez} {et~al.}(2019){Mart{\'\i}nez
  Gonz{\'a}lez}, {Gonz{\'a}lez-Fern{\'a}ndez}, {Asensio Ramos},
  {Socas-Navarro}, {Westendorp Plaza}, {Boyajian}, {Wright}, {Collier Cameron},
  {Gonz{\'a}lez Hern{\'a}ndez}, {Holgado}, {Kennedy}, {Masseron}, {Molinari},
  {Saario}, {Sim{\'o}n-D{\'\i}az}, \&
  {Toledo-Padr{\'o}n}}]{martinez_highres_notzotero}
{Mart{\'\i}nez Gonz{\'a}lez}, M.~J., {Gonz{\'a}lez-Fern{\'a}ndez}, C., {Asensio
  Ramos}, A., {et~al.} 2019, \mnras, 486, 236

\bibitem[{{Metchev} \& {Hillenbrand}(2009)}]{Metchev2009}
{Metchev}, S.~A., \& {Hillenbrand}, L.~A. 2009, \apjs, 181, 62

\bibitem[{Montet \& Simon(2016)}]{montet_kic_2016}
Montet, B.~T., \& Simon, J.~D. 2016, The Astrophysical Journal Letters, 830,
  L39.
\newblock \url{http://adsabs.harvard.edu/abs/2016ApJ...830L..39M}

\bibitem[{{Pearce} {et~al.}(2019){Pearce}, {Kraus}, {Dupuy}, {Ireland },
  {Rizzuto}, {Bowler}, {Birchall}, \& {Wallace}}]{Pearce2019GSC6214-210}
{Pearce}, L.~A., {Kraus}, A.~L., {Dupuy}, T.~J., {et~al.} 2019, \aj, 157, 71

\bibitem[{{Pecaut} \& {Mamajek}(2016)}]{PecautMamajek2016}
{Pecaut}, M.~J., \& {Mamajek}, E.~E. 2016, \mnras, 461, 794

\bibitem[{{Price-Whelan} {et~al.}(2018){Price-Whelan}, {Sip{'{o}}cz},
  {G{"u}nther}, {Lim}, {Crawford}, {Conseil}, {Shupe}, {Craig}, {Dencheva},
  {Ginsburg}, {VanderPlas}, {Bradley}, {P{'e}rez-Su{'a}rez}, {de Val-Borro},
  {Paper Contributors}, {Aldcroft}, {Cruz}, {Robitaille}, {Tollerud},
  {Coordination Committee}, {Ardelean}, {Babej}, {Bach}, {Bachetti}, {Bakanov},
  {Bamford}, {Barentsen}, {Barmby}, {Baumbach}, {Berry}, {Biscani}, {Boquien},
  {Bostroem}, {Bouma}, {Brammer}, {Bray}, {Breytenbach}, {Buddelmeijer},
  {Burke}, {Calderone}, {Cano Rodr{'i}guez}, {Cara}, {Cardoso}, {Cheedella},
  {Copin}, {Corrales}, {Crichton}, {D{ extquoteright}Avella}, {Deil},
  {Depagne}, {Dietrich}, {Donath}, {Droettboom}, {Earl}, {Erben}, {Fabbro},
  {Ferreira}, {Finethy}, {Fox}, {Garrison}, {Gibbons}, {Goldstein}, {Gommers},
  {Greco}, {Greenfield}, {Groener}, {Grollier}, {Hagen}, {Hirst}, {Homeier},
  {Horton}, {Hosseinzadeh}, {Hu}, {Hunkeler}, {Ivezi{'c}}, {Jain}, {Jenness},
  {Kanarek}, {Kendrew}, {Kern}, {Kerzendorf}, {Khvalko}, {King}, {Kirkby},
  {Kulkarni}, {Kumar}, {Lee}, {Lenz}, {Littlefair}, {Ma}, {Macleod},
  {Mastropietro}, {McCully}, {Montagnac}, {Morris}, {Mueller}, {Mumford},
  {Muna}, {Murphy}, {Nelson}, {Nguyen}, {Ninan}, {N{"o}the}, {Ogaz}, {Oh},
  {Parejko}, {Parley}, {Pascual}, {Patil}, {Patil}, {Plunkett}, {Prochaska},
  {Rastogi}, {Reddy Janga}, {Sabater}, {Sakurikar}, {Seifert}, {Sherbert},
  {Sherwood-Taylor}, {Shih}, {Sick}, {Silbiger}, {Singanamalla}, {Singer},
  {Sladen}, {Sooley}, {Sornarajah}, {Streicher}, {Teuben}, {Thomas},
  {Tremblay}, {Turner}, {Terr{'o}n}, {van Kerkwijk}, {de la Vega}, {Watkins},
  {Weaver}, {Whitmore}, {Woillez}, {Zabalza}, \& {Contributors}}]{astropy:2018}
{Price-Whelan}, A.~M., {Sip{'{o}}cz}, B.~M., {G{"u}nther}, H.~M., {et~al.}
  2018, aj, 156, 123

\bibitem[{Raghavan {et~al.}(2010)Raghavan, McAlister, Henry, Latham, Marcy,
  Mason, Gies, White, \& ten Brummelaar}]{raghavan_survey_2010}
Raghavan, D., McAlister, H.~A., Henry, T.~J., {et~al.} 2010, The Astrophysical
  Journal Supplement Series, 190, 1.
\newblock \url{http://adsabs.harvard.edu/abs/2010ApJS..190....1R}

\bibitem[{Service {et~al.}(2016)Service, Lu, Campbell, Sitarski, Ghez, \&
  Anderson}]{service_new_2016}
Service, M., Lu, J.~R., Campbell, R., {et~al.} 2016, Publications of the
  Astronomical Society of the Pacific, 128, 095004.
\newblock \url{http://adsabs.harvard.edu/abs/2016PASP..128i5004S}

\bibitem[{Simon {et~al.}(2018)Simon, Shappee, Pojmański, Montet, Kochanek, van
  Saders, Holoien, \& Henden}]{simon_where_2018}
Simon, J.~D., Shappee, B.~J., Pojmański, G., {et~al.} 2018, The Astrophysical
  Journal, 853, 77.
\newblock \url{http://adsabs.harvard.edu/abs/2018ApJ...853...77S}

\bibitem[{Skrutskie {et~al.}(2006)Skrutskie, Cutri, Stiening, Weinberg,
  Schneider, Carpenter, Beichman, Capps, Chester, Elias, Huchra, Liebert,
  Lonsdale, Monet, Price, Seitzer, Jarrett, Kirkpatrick, Gizis, Howard, Evans,
  Fowler, Fullmer, Hurt, Light, Kopan, Marsh, McCallon, Tam, Van~Dyk, \&
  Wheelock}]{skrutskie_two_2006}
Skrutskie, M.~F., Cutri, R.~M., Stiening, R., {et~al.} 2006, The Astronomical
  Journal, 131, 1163.
\newblock \url{http://adsabs.harvard.edu/abs/2006AJ....131.1163S}

\bibitem[{Thompson {et~al.}(2016)Thompson, Scicluna, Kemper, Geach, Dunham,
  Morata, Ertel, Ho, Dempsey, Coulson, Petitpas, \&
  Kristensen}]{thompson_constraints_2016}
Thompson, M.~A., Scicluna, P., Kemper, F., {et~al.} 2016, Monthly Notices of
  the Royal Astronomical Society, 458, L39.
\newblock \url{http://adsabs.harvard.edu/abs/2016MNRAS.458L..39T}

\bibitem[{Virtanen {et~al.}(2020)Virtanen, Gommers, Oliphant, Haberland, Reddy,
  Cournapeau, Burovski, Peterson, Weckesser, Bright, van~der Walt, Brett,
  Wilson, Millman, Mayorov, Nelson, Jones, Kern, Larson, Carey, Polat, Feng,
  Moore, VanderPlas, Laxalde, Perktold, Cimrman, Henriksen, Quintero, Harris,
  Archibald, Ribeiro, Pedregosa, \& van Mulbregt}]{virtanen_scipy_2020}
Virtanen, P., Gommers, R., Oliphant, T.~E., {et~al.} 2020, Nature Methods, 17,
  261, number: 3 Publisher: Nature Publishing Group.
\newblock \url{https://www.nature.com/articles/s41592-019-0686-2}

\bibitem[{{Wizinowich} {et~al.}(2000){Wizinowich}, {Acton}, {Shelton},
  {Stomski}, {Gathright}, {Ho}, {Lupton}, {Tsubota}, {Lai}, {Max}, {Brase},
  {An}, {Avicola}, {Olivier}, {Gavel}, {Macintosh}, {Ghez}, \&
  {Larkin}}]{Wizinowich2000}
{Wizinowich}, P., {Acton}, D.~S., {Shelton}, C., {et~al.} 2000, Publications of
  the Astronomical Society of the Pacific, 112, 315

\bibitem[{Wright {et~al.}(2016)Wright, Cartier, Zhao, Jontof-Hutter, \&
  Ford}]{wright_search_2016}
Wright, J.~T., Cartier, K. M.~S., Zhao, M., Jontof-Hutter, D., \& Ford, E.~B.
  2016, The Astrophysical Journal, 816, 17.
\newblock \url{http://adsabs.harvard.edu/abs/2016ApJ...816...17W}

\bibitem[{Wright \& Sigurdsson(2016)}]{wright_families_2016}
Wright, J.~T., \& Sigurdsson, S. 2016, The Astrophysical Journal Letters, 829,
  L3.
\newblock \url{http://adsabs.harvard.edu/abs/2016ApJ...829L...3W}

\bibitem[{Wyatt {et~al.}(2018)Wyatt, van Lieshout, Kennedy, \&
  Boyajian}]{wyatt_modelling_2018}
Wyatt, M.~C., van Lieshout, R., Kennedy, G.~M., \& Boyajian, T.~S. 2018,
  Monthly Notices of the Royal Astronomical Society, 473, 5286.
\newblock \url{http://adsabs.harvard.edu/abs/2018MNRAS.473.5286W}

\bibitem[{Yelda {et~al.}(2010)Yelda, Lu, Ghez, Clarkson, Anderson, Do, \&
  Matthews}]{yelda_improving_2010}
Yelda, S., Lu, J.~R., Ghez, A.~M., {et~al.} 2010, The Astrophysical Journal,
  725, 331.
\newblock \url{http://adsabs.harvard.edu/abs/2010ApJ...725..331Y}

\bibitem[{{Zakamska} \& {Tremaine}(2004)}]{Zakamska2004_excitiation}
{Zakamska}, N.~L., \& {Tremaine}, S. 2004, \aj, 128, 869

\bibitem[{{Zinn} {et~al.}(2019){Zinn}, {Pinsonneault}, {Huber}, \&
  {Stello}}]{zinn19}
{Zinn}, J.~C., {Pinsonneault}, M.~H., {Huber}, D., \& {Stello}, D. 2019, \apj,
  878, 136

\end{thebibliography}

\end{document}